\documentclass[iop]{emulateapj}

\usepackage{graphicx}
\usepackage{amsmath}
\usepackage{amssymb}
\usepackage[T1]{fontenc}
\usepackage{color}
\usepackage{times}
\usepackage{multirow}

\newcommand{\jcap}{J.~of Cos.~and Astropart.~Phys.}
\newcommand{\tabh}{\parbox[0pt][1.9em][c]{0cm}{}}

\newcommand{\tab}{\parbox[0pt][1.5em][c]{0cm}{}}

\begin{document}

\title[X-Ray Emission from Star-Forming Galaxies]{X-Ray Emission from Star-Forming Galaxies \\ - Signatures of Cosmic Rays and Magnetic Fields}



\author{Jennifer Schober\altaffilmark{1},
        Dominik R.~G.~Schleicher\altaffilmark{2}, 
        Ralf S.~Klessen\altaffilmark{1}}
\email{schober@stud.uni-heidelberg.de}
\altaffiltext{1}{Zentrum f\"ur Astronomie der Universit\"at Heidelberg, Institut f\"ur Theoretische Astrophysik, Albert-Ueberle-Str.~2, D-69120 Heidelberg, Germany}
\altaffiltext{2}{Georg-August-Universit\"at G\"ottingen, Institut f\"ur Astrophysik, Friedrich-Hund-Platz, 37077 G\"ottingen, Germany}

\date{\today}

\begin{abstract}
The evolution of magnetic fields in galaxies is still an open problem in astrophysics. In nearby galaxies the far-infrared-radio correlation indicates the coupling between magnetic fields and star formation. The correlation arises from the synchrotron emission of cosmic ray electrons traveling through the interstellar magnetic fields. However, with an increase of the interstellar radiation field (ISRF), inverse Compton scattering becomes the dominant energy loss mechanism of cosmic ray electrons with a typical emission frequency in the X-ray regime. The ISRF depends on the one hand on the star formation rate and becomes stronger in starburst galaxies, and on the other hand increases with redshift due to the evolution of the cosmic microwave background. With a model for the star formation rate of galaxies, the ISRF, and the cosmic ray spectrum, we can calculate the expected X-ray luminosity resulting from the inverse Compton emission. Except for galaxies with an active galactic nucleus the main additional contribution to the X-ray luminosity comes from X-ray binaries. We estimate this contribution with an analytical model as well as with an observational relation, and compare it to the pure inverse Compton luminosity. Using data from the \textit{Chandra} Deep Field Survey and far-infrared observations from ALMA we then determine upper limits for the cosmic ray energy. Assuming that the magnetic energy in a galaxy is in equipartition with the energy density of the cosmic rays, we obtain upper limits for the magnetic field strength. Our results suggest that the mean magnetic energy of young galaxies is similar to the one in local galaxies. This points toward an early generation of galactic magnetic fields, which is in agreement with current dynamo evolution models.
\end{abstract}


\maketitle

\section{Introduction}

Observations show that magnetic fields contribute significantly to a galaxys energy budget. The current picture predicts the magnetic energy density to be roughly $0.89~\mathrm{erg~cm}^{-3}$, which is comparable to the thermal kinetic energy density with roughly $0.49~\mathrm{erg~cm}^{-3}$, and the energy density of cosmic rays to be $1.39~\mathrm{erg~cm}^{-3}$ \citep{Draine1978}. Moreover, the magnetic energy is distributed over many orders of magnitudes in physical length scales. It is thus expected that the magnetic field plays a major role in the dynamics of the whole galaxy and also on smaller scales down to individual star formation processes. \\ 
The structure of magnetic fields in local galaxies is known quiet well \citep{BeckEtAl1999,Beck2011}. A spiral galaxy typically shows a large-scale magnetic field, which follows the optical spiral arms and is strongest in the interarm regions. The typical coherence length is 10 kpc and the strength roughly $10^{-5}$ G. Even more important in terms of the energy density is the small-scale unordered magnetic field, which exceeds the one of the ordered field by a factor of a few. \\
The origin and evolution of galactic magnetic fields is still an active field of research with many open questions to answer \citep{KulsrudZweibel2008}. Theory predicts that unordered fields were generated already in young galaxies by a turbulent dynamo. This mechanism amplifies weak magnetic seed fields by randomly stretching, twisting, and folding the field lines in turbulent motions \citep{Kazantsev1968,BrandenburgSubramanian2005,SchoberEtAl2012.1,SchoberEtAl2012.3,BovinoEtAl2013}. Semi-analytical calculations \citep{SchoberEtAl2013} as well as numerical simulations \citep{BeckEtAl2012,LatifEtAl2013} show that the turbulent dynamo can produce a field of the order of $10^{-6}$ G within a few Myrs. The large-scale magnetic field is likely produced by a large-scale galactic dynamo, which operates on much longer timescales then the turbulent dynamo. \\
In order to test the evolution scenario of galactic fields, in addition to the analytical and numerical calculations an observational test is essential. However, the problem is that standard methods for magnetic field observations are difficult to pursue at high redshifts. Only indirect observations like the CMB bispectrum \citep{ShiraishiEtAl2012}, the non-detection of TeV blazers \citep{NeronovVovk2010} and Faraday rotation measurements \citep{HammondRobishawGaensler2012}, which detect the magnetic field strength along the line of sight, can be applied at high redshifts. \\
A very frequently used method to estimate magnetic field strengths in galaxies is synchrotron emission, which is observed in the radio band. This type of radiation is emitted by high energy cosmic ray electrons traveling through the magnetized interstellar medium (ISM). With the intensity of synchrotron emission one can calculate the energy density of cosmic rays. By assuming that cosmic rays and interstellar magnetic fields are in energy equilibrium the magnetic field strength can be computed \citep{BeckKrause2005}. \\
A further important observation was made by \cite{YunEtAl2001}, who observed a correlation between the radio flux and the far-infrared (FIR) flux. This FIR-radio correlation shows a coupling between the star formation rate (SFR), which determines the FIR flux, and the magnetic field in the ISM. A new interpretation of this correlation was suggested by \cite{SchleicherBeck2013}. They claim that the supernova rate, which is proportional to the SFR, sets the amount of turbulence in the ISM, which in turn determines the magnetic energy produced by turbulent dynamo. Due to energy conservation and additional efficiency effects a turbulent dynamo can only convert a certain fraction of turbulent kinetic energy into magnetic energy \citep{FederrathEtAl2011.2}. A coupling between the SFR (FIR flux) and the magnetic field (radio flux) can thus be assumed in local galaxies. \\
But what happens in higher redshifted galaxies? Here one needs to take into account the rapidly growing number of cosmic microwave background (CMB) photons. These can interact with the cosmic ray electrons in inverse Compton scattering, typically resulting in X-ray photons. \cite{SchleicherBeck2013} have shown that inverse Compton scattering is in fact the dominant energy loss mechanism of cosmic ray electrons at high redshifts. Thus, we expect a breakdown of the FIR-radio correlation and X-ray bright galaxies above a critical redshift. \\
We propose here a method based on the inverse Compton scattering process to gain information about cosmic rays and magnetic fields in young galaxies. With a given SFR and interstellar radiation field (ISRF) we determine the inverse Compton component of the X-ray luminosity of a redshifted star-forming galaxy. From this we calculate the energy of the cosmic ray electrons and the resulting total cosmic ray energy density. By assuming equipartition between the cosmic ray energy density and the magnetic field energy density, we are able to predict an upper limit of the field strength. \\
New instruments provide exceptionally good data of galaxies at very high redshifts. Especially the deep fields of the \textit{Chandra} satellite\footnote{http://chandra.harvard.edu/}, the extended \textit{Chandra} Deep Field-South (E-CDF-S) and the \textit{Chandra} Deep Field-North (CDF-N), include lots of information about the X-ray properties of extremely low luminosity objects. As a very important future tool we discuss also limits that will be obtained by X-ray observatory \textit{Athena+}\footnote{http://www.the-athena-x-ray-observatory.eu/}. Combination with the new FIR data from the Atacama Large Millimeter/submillimeter Array (ALMA\footnote{http://www.almaobservatory.org/}) can lead to new conclusions. The ALMA LABOCA E-CDF-S Submillimeter Survey makes a multi-wavelength analyses possible. \\
The paper is structured as follows: We present our model of young galaxies in section \ref{Galaxies}, including the SFR, the ISRF and the cosmic ray spectrum. In section \ref{Breakdown} we summarize the results of \cite{SchleicherBeck2013}, who proposed the breakdown of the FIR-radio correlation. The combination of our ISRF and the cosmic ray spectrum results in a typical inverse Compton spectrum. The derivation of the inverse Compton X-ray luminosity is given in section \ref{ExpectedIC}. We discuss additional X-ray sources in section \ref{OtherXrays}. In the last section, section \ref{Examples}, we apply our model to some exemplary galaxies, for which data from \textit{Chandra} and ALMA is available. We draw our conclusions in section \ref{Conclusions}.

\newpage



\section{Model of Typical Galaxies}
\label{Galaxies}


For exploring the X-ray properties of galaxies we use two different models: a model for a ``normal'' galaxy and one for a starburst galaxy. The models differ obviously in their star formation rates, which has important consequences for the interstellar radiation field and the number density of cosmic rays. We report the details of our models in the following.

\subsection{General Aspects}

\begin{figure}
  \includegraphics[width=0.5\textwidth]{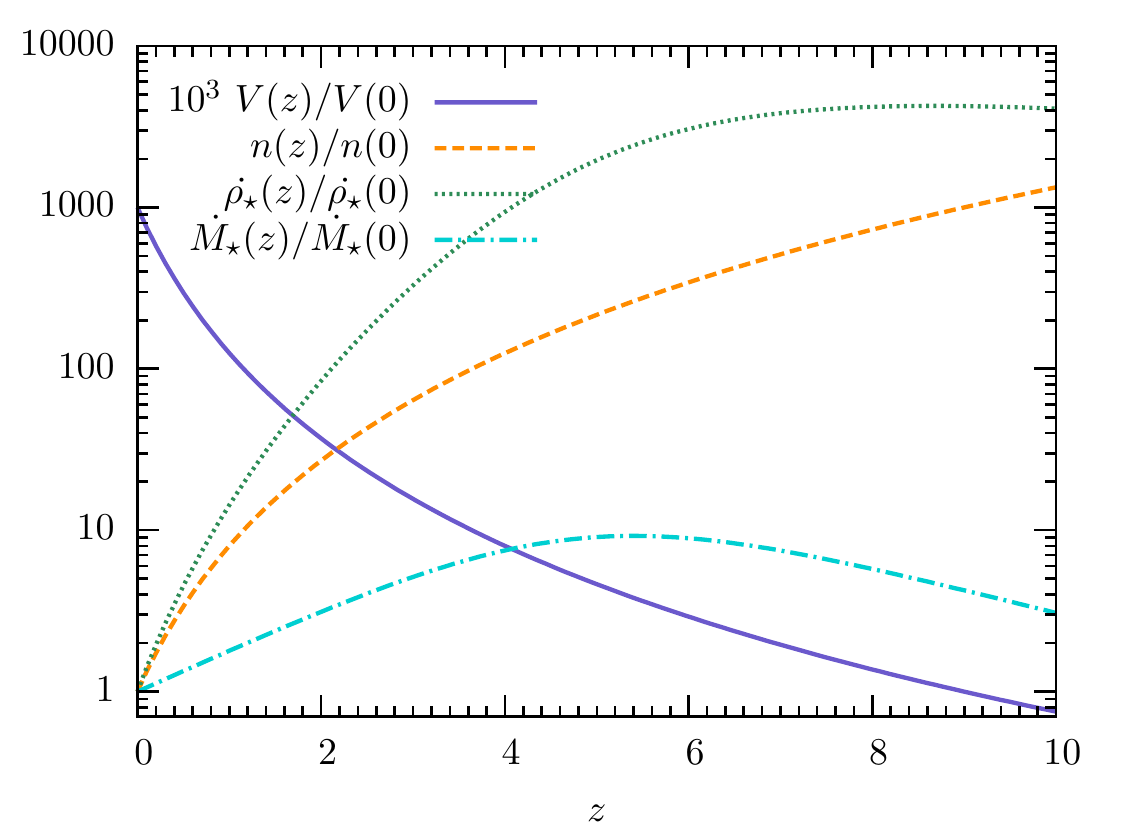}
  \caption{The evolution of the normalized star formation rate density $\dot{\rho_\star}(z)/\dot{\rho_\star}(0)$ \citep{HernquistSpringel2003} and the star formation rate $\dot{M_\star}(z)/\dot{M_\star}(0)$ as a function of redshift $z$. We also show the evolution of the normalized volume of the galaxy $V(z)/V(0)$, which has been multiplied by a factor of $10^3$ for better visualization, and the normalized particle density $n(z)/n(0)$. \vspace{0.3cm}}
\label{VgalSFR_z}
\end{figure}

We use a geometrically very simple model of a galaxy, which has the shape of a disk. The radius $R(z)$ and the scale height $H(z)$ evolve with redshift $z$. \citet{FergusonEtAl2004} find a change of radius proportional to $(1+z)^{-1}$ for a fixed circular velocity and proportional to $(1+z)^{-2/3}$ for a fixed mass. Observations of galaxy evolution show that the mean scaling of the galaxy size lies in between these two extrema. We choose the former model and a assume
\begin{equation}
  R(z) = R_0 (1+z)^{-1}
\end{equation}
and a scale height of
\begin{equation}
  H(z) = H_0 (1+z)^{-1},
\end{equation}
leading to a galaxy volume of 
\begin{equation}
  V(z) = \pi R_0^2 H_0 (1+z)^{-3}.
\label{Vgal}
\end{equation}
The normalization of the radius and the scale height, $R_0$ and $H_0$, are set by the galaxies at present day. We analyze two types of galaxies: a normal Milky Way like galaxy, for which we use $R_0 = 1.5\times10^4~\mathrm{pc}$ and $H_0 = 500~\mathrm{pc}$ \citep{Ferriere2001}, and a starburst galaxy, which is of a similar type as M82. The radius of the central starburst region is roughly $R_0 = 300~\mathrm{pc}$ with a scale height of $R_0 = 200~\mathrm{pc}$ \citep{deCeadelPozoEtAl2009}. We note, that our model of perfect disks is very idealized as the scale height usually changes with the radius. \\
Further, we assume the particle density of the ISM to scale as
\begin{equation}
  n(z) = n_0 (1+z)^3.
\label{n}
\end{equation}
Our fiducial values for the present-day density $n_0$ are listed in table \ref{Table_Props}. With our model of a uniform density we simplify real galaxies, where there are gradients in density. \\
The evolution of the normalized density $n(z)/n(0)$ and galaxy volume $V(z)/V(0)$, which are proportional to $(1+z)^{3}$ or $(1+z)^{-3}$, respectively, are shown in figure \ref{VgalSFR_z}. Note, that our model depends initially only on the normalized volume. Only when applied to real data in section \ref{Examples}, we need the volume of the galaxy when calculating the magnetic field strength from the total magnetic energy. 
\begin{table} 
     \begin{tabular}{lll}
      \hline  \hline  
      \tabh     ~ 							& normal galaxy		&  starburst core  \\
      \hline
      \tab  $R_0$~[pc] 							& $1.5 \times 10^4$  	&  $300$	\\
      \tab  $H_0$~[pc] 							& $500$  		&  $200$	\\
      \tab  $n_0$~[cm$^{-3}$] 						& $3$ 			&  $300$ 	\\
      \tab  $\dot{M}_\star(0)$~[M$_\odot$~yr$^{-1}$]     		& $2$   		&  $10$   	\\
      \tab  $\dot{\rho}_\star(0)$~[M$_\odot$~yr$^{-1}$~pc$^{-3}$]  	& $1.9 \times 10^{-12}$ &  $3.5 \times 10^{-8}$ \\
      \hline  \hline  
    \end{tabular}
    \caption{Properties of our two fiducial models: a normal galaxy comparable to the Milky Way and a starburst galaxy comparable to M82.}
  \label{Table_Props}
\end{table}

\newpage

\subsection{Star Formation Rate}

\subsubsection{Model for the Star Formation History}
In section \ref{ExpectedIC} we look in the X-ray evolution of an idealized galaxy, which evolves according to the cosmic mean star formation history \citep{MadauEtAl1996,MadauEtAl1998,SteidelEtAl1999,MadauPozzetti2000}. The mean star formation rate of the Universe has been studied in simulations by \cite{HernquistSpringel2003}. We use their relation to follow the evolution of a characteristic galaxy. With this model we get an idea of the inverse Compton scattering process as a function of redshift.\\
The star formation rate (SFR) of a galaxy is defined as
\begin{equation}
  \dot{M}_{\star}(z) = V(z)~\dot{\rho}_{\star}(z),
\label{SFR}
\end{equation}
with the volume of the galaxy (\ref{Vgal}) and the star formation rate density \citep{HernquistSpringel2003}
\begin{equation}
  \dot{\rho}_{\star}(z) \propto \frac{\kappa_2~\mathrm{exp}[\kappa_1(z-z_\mathrm{m})]}{\kappa_2-\kappa_1+\kappa_1~\mathrm{exp}[\kappa_2~(z-z_\mathrm{m})]} (1+z)^3,
\label{SFRdensity}
\end{equation}
with the parameters $\kappa_1 = 3/5$, $\kappa_2 = 14/15$ and $z_\mathrm{m} = 5.4$. The factor $(1+z)^3$ in equation (\ref{SFRdensity}) comes from the conversion from comoving into physical units.\\
In figure \ref{VgalSFR_z} we show the evolution of the normalized star formation rate density $\dot{\rho}_{\star}(z)/\dot{\rho}_{\star}(0)$ and the normalized star formation rate $\dot{M}_{\star}(z)/\dot{M}_{\star}(0)$ with redshift. For the normalization of the star formation rate we use a typical value observed in the Milky Way $\dot{M}_{\star}(0)=2~\mathrm{\mathrm{M}_\odot~yr}^{-1}$ for the model of a normal galaxy. Note, that the proposals of the galactic SFR differ widely. While for example \citet{DiehlEtAl2006} find a value of $4~\mathrm{\mathrm{M}_\odot~yr}^{-1}$ from gamma ray observations, \textit{Spitzer} observations suggest that the SFR is as low as $0.68-1.45~\mathrm{\mathrm{M}_\odot~yr}^{-1}$ \citep{RobitailleWhitney2010}. For the starburst model we use $\dot{M}_{\star}(0)=10~\mathrm{\mathrm{M}_\odot~yr}^{-1}$, which is close to the observed value of the starburst galaxy M82 \citep{FoersterSchreiberEtAl2003}. 
\begin{figure*}
  \includegraphics[width=\textwidth]{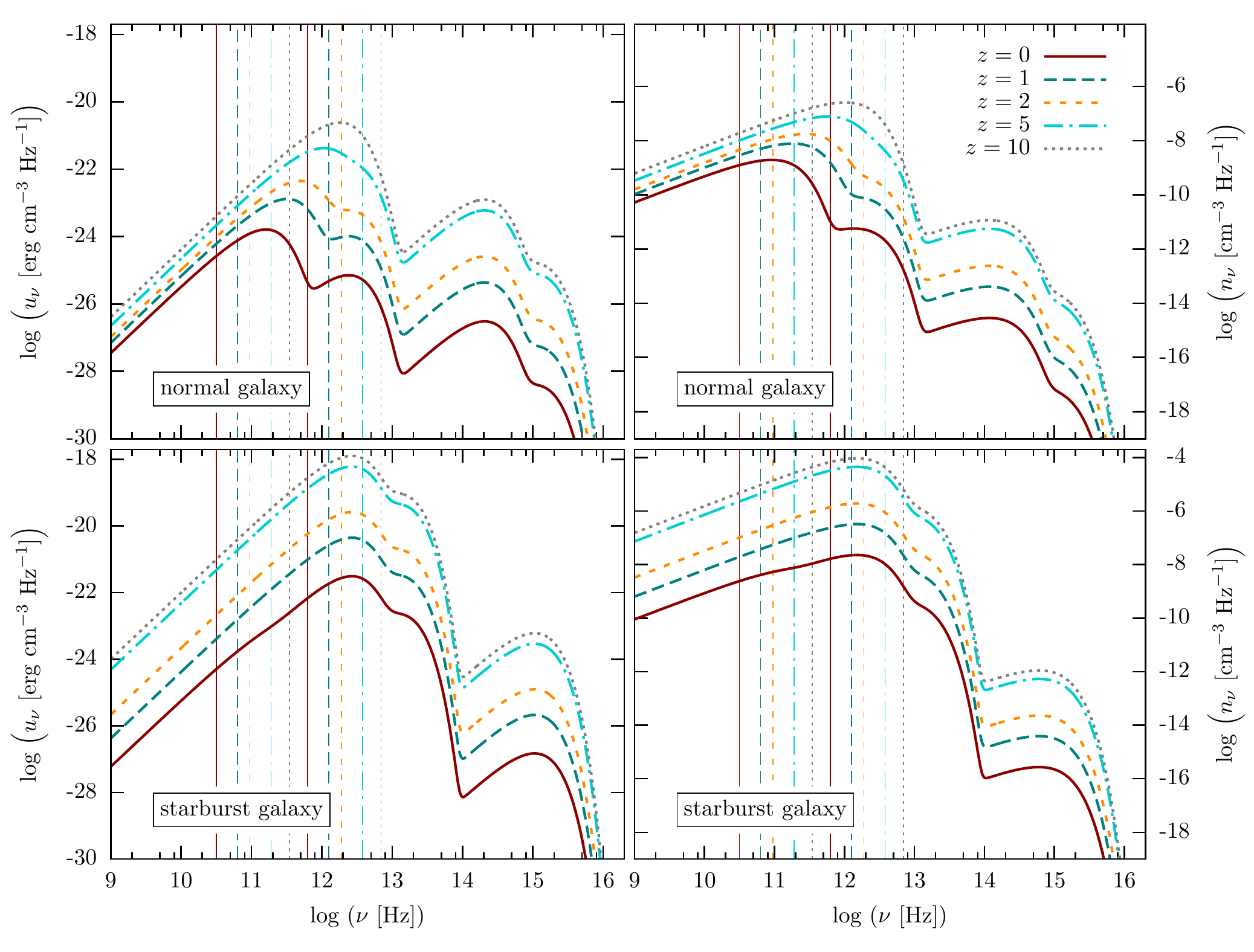}
  \caption{The spectral energy density $u_{\mathrm{ISRF},\nu}$ (left hand side) and the spectral photon distribution $n_{\mathrm{ISRF},\nu}$ (right hand side) of the interstellar radiation field for different redshifts between $z=0-10$. In the top panels we show our model for galaxies with a normal star formation rate and in the lower panels for a starburst galaxy. The parameters of the model are summarized in table \ref{Table_ISRF}. The vertical lines indicate typical frequency range of photons that are inverse Compton scattered into the X-ray regime (see section \ref{ExpectedIC_CharacFreq} for more details). \vspace{0.19 cm}}
\label{ISRF_nu}
\end{figure*} 

\newpage

\subsubsection{Observed Star Formation Rate}
The star formation history of an arbitrary galaxy will, however, differ significantly from this idealized picture. There will be phases of extremely high SFR induced, for example, by mergers with other galaxies followed by quiet phases. Thus, for getting information about the detailed emission processes of a single galaxy it is better to use direct observational input of the SFR. We use the model of the star formation history only in places, where we discuss the general trend of galaxy evolution, while we take a fixed SFR for single galaxies.\\
The SFR of a galaxy can be estimated from the observed infrared flux. \citet{Kennicutt1998} finds the following correlation between the SFR $\dot{M}_\star$ and the infrared luminosity $L_\mathrm{IR}$:
\begin{equation}
  \dot{M}_\star(z) = 1.8\times10^{-10}~\mathrm{\mathrm{M}_\odot~yr}^{-1}~\left(\frac{L_\mathrm{IR}}{\mathrm{L}_\odot}\right),
\label{SFR_LIR}
\end{equation}
where $\mathrm{L}_\odot \approx 3.8 \times10^{33}~\mathrm{erg~s}^{-1}$ is the solar luminosity. In their data set \citet{WangEtAl2013} employ NIR-through-radio SED fitting according to the work of \citet{SwinbankEtAl2013} to calculate $L_\mathrm{IR}$. We will use examples of this multi-wavelength data set in section \ref{Examples}.

\subsection{Supernova Rate}

Cosmic rays are believed to origin in supernova shock fronts. Thus, for determination of the number and energy of the cosmic rays in a galaxy, the supernova rate is an important input. \\
We assume here a Kroupa initial mass function of stars, which decreases proportional as the stellar mass to the power of -1.3 in the range of 0.08 to 0.5 $\mathrm{M}_\odot$ and to the power of -2.3 for larger SFRs \citep{Kroupa2002}. The number of supernova per time is then roughly
\begin{eqnarray}
  \dot{N}_\mathrm{SN} = 0.156~\frac{\dot{M}_\star}{12.26~\mathrm{M}_\odot},
\label{SNrate}
\end{eqnarray}
with $12.26~\mathrm{M}_\odot$ being the mean mass of a supernova candidate.

\subsection{Interstellar Radiation Field}

An essential role for the inverse Compton scattering plays the ISRF with which the cosmic rays interact. In our model we consider five different radiation components: the thermal (cold and warm) infrared (IR), optical (opt), ultraviolet (UV) radiation and the cosmic microwave background (CMB) \citep{CirelliPanci2009,ChakrabortyFields2013}. The interstellar radiation field (ISRF) can then be approximated by the sum of the individual Planck spectra,
\begin{equation}
  u_{\mathrm{ISRF},\nu} = \sum_i f_i~\frac{8 \pi h}{c^3}~\frac{\nu^3}{\mathrm{exp}(h\nu/(k T_i) - 1},
\label{uISRF}
\end{equation}
with $i = \mathrm{IR,opt,UV,CMB}$. The dimensionless weights $f_i$ as well as the different temperatures $T_i$ at $z=0$ are taken from \cite{ChakrabortyFields2013}. Note, that we model a redshift dependence, which is given in table \ref{Table_ISRF}. For this we use $T_\mathrm{CMB}(z) = T_\mathrm{CMB}(0)~(1+z)$ and multiply the weights $f_\mathrm{IR}$, $f_\mathrm{opt}$, $f_\mathrm{UV}$ with the normalized star formation rate density $\dot{\rho}_\star(z)/\dot{\rho}_\star(0)$. The resulting spectral energy density $u_{\mathrm{ISRF},\nu}$ is shown in the left panel of figure \ref{ISRF_nu}. \\
\begin{table*} 
     \begin{tabular}{lllllll}
      \hline  \hline  
      \tabh     ~ & ~ 		     & UV 		   & optical 	    &  IR (warm) & IR (cold) 	      & CMB           \\
      \cline{2-7}
      \tab normal:    & $f_i$  	     & $8.4\times10^{-17} \frac{\dot{\rho}_\star(z)}{\dot{\rho}_{\star,\mathrm{MW}}}$ & $8.9\times10^{-13} \frac{\dot{\rho}_\star(z)}{\dot{\rho}_{\star,\mathrm{MW}}}$ &	- 	 & $1.3\times10^{-5} \frac{\dot{\rho}_\star(z)}{\dot{\rho}_{\star,\mathrm{MW}}}$ &  1           \\
      \tab  ~         & $T_i$ [K] ~  & $1.8\times10^{4}$   & $3.5\times10^{3}$   & 	-	 & $41$  	      & $2.73(1+z)$        \\
      \cline{2-7}
      \tab starburst: & $f_i$        & $3.2\times10^{-15} \frac{\dot{\rho}_\star(z)}{\dot{\rho}_{\star,\mathrm{M82}}}$ & $0.0$		    & $3.61\times10^{-5}\frac{\dot{\rho}_\star(z)}{\dot{\rho}_{\star,\mathrm{M82}}}$ & $4.22\times10^{-2}\frac{\dot{\rho}_\star(z)}{\dot{\rho}_{\star,\mathrm{M82}}}$  & 1   \\
      \tab ~          & $T_i$ [K] ~~ & $1.8\times10^{4}$   &	$3.5\times10^{3}$   &	$200$	 & $45$       	      & $2.73 (1+z)$ \\
      \hline  \hline  
    \end{tabular}
    \caption{A model of the interstellar radiation field, which includes five different radiation components: ultraviolet (UV) radiation, optical radiation, thermal (warm and cold) infrared (IR) radiation and the cosmic microwave background (CMB) (see \cite{CirelliPanci2009} and \cite{ChakrabortyFields2013}). We give here the dimensionless weights compared to the CMB $f_i$, which include a scaling with the normalized star formation rate density $\dot{\rho}_\star(z)/\dot{\rho}_{\star,\mathrm{MW}}$ or $\dot{\rho}_\star(z)/\dot{\rho}_{\star,\mathrm{M82}}$, respectively, and the temperatures $T_i$.  \vspace{0.2 cm}} 
  \label{Table_ISRF}
\end{table*}
The total energy density of the interstellar radiation field is
\begin{eqnarray}
  u_{\mathrm{ISRF}}  =  \int_0^\infty u_{\mathrm{ISRF},\nu}~\mathrm{d}\nu = \frac{8~\pi^5 k^4}{15~c^3 h^3}~\sum_i f_i T_i^4
\label{uISRFtot}
\end{eqnarray}
From the energy spectrum (\ref{uISRF}) we can calculate the photon distribution by
\begin{equation}
  n_{\mathrm{ISRF},\nu} = \frac{u_{\mathrm{ISRF},\nu}}{h\nu}.
\label{nISRF}
\end{equation}
The result is shown in right panel of figure \ref{ISRF_nu}. Note, that as well $u_{\mathrm{ISRF}}$ as $n_{\mathrm{ISRF}}$ increase with redshift, as the SFR density and the CMB density constantly increase with $z$. The peak of the CMB component further moves to higher frequencies with $z$ due to the increasing CMB temperature.

\subsection{Cosmic Rays}
\label{CR_Model}

The origin of high energy cosmic rays is commonly believed to be first-order Fermi shock acceleration in supernova remnants and extragalactic sources \citep{Bell1978a,Bell1978b,Drury1983,Schlickeiser2002}. However, there are additional models like acceleration by MHD waves \citep{SchlickeiserMiller1998, BrunettiEtAl2001, FujitaEtAl2003} and magnetic reconnection \citep{deGouveiadalPinoLazarian2005}. All these theoretical models result in a power-law distribution. \\
The injection spectrum of cosmic ray protons can be described by 
\begin{equation}
  Q_{\mathrm{p}}(\gamma_{\mathrm{p}}) = Q_{\mathrm{p},0}~\gamma_\mathrm{p}^{-\chi},
\label{Qp}
\end{equation}
with the Lorentz factor of protons $\gamma_\mathrm{p}$. The exponent $\chi$ depends strongly on the properties of the cosmic ray acceleration site, i.e.~the supernova shock front. First order Fermi acceleration theory predicts for strong shocks a value of $\chi=2.0$ for non-relativistic gas and $\chi=2.5$ for a relativistic gas \citep{Bell1978b}. More detailed models of supernova shock fronts \citep{BogdanVolk1983} result in $\chi = 2.1 - 2.3$. We will use here a fiducial value of $\chi = 2.2$. \\
We normalize the proton injection spectrum with a total energy injection rate $\xi E_\mathrm{SN} \dot{N}_\mathrm{SN}$, where $E_\mathrm{SN}$ is the energy of one supernova and $\dot{N}_\mathrm{SN}$ the supernova rate (\ref{SNrate}). This yields a proportionality factor in (\ref{Qp}) of
\begin{equation}
  Q_{\mathrm{p},0} = \frac{\xi E_\mathrm{SN} \dot{N}_\mathrm{SN} (\chi-2)}{m_\mathrm{p} c^2 ~ \gamma_\mathrm{p,0}^{2-\chi}}.
\label{Qp0}
\end{equation}
A typical value of $\xi$, which is the fraction of the total energy released in supernovae that goes into cosmic rays, is given in the literature as 0.1. We use this as our fiducial value. Moreover, we will analyse values from $\xi=0.05$ to $\xi=0.2$, as simulations suggest that there is a density dependency of $\xi$ \citep{Dorfi2000}. As the upper end of the cosmic ray energy spectrum, which extends up to $10^{21}$ eV per particle, does not contribute significantly to the total cosmic ray energy, only the lower end of the spectrum $\gamma_\mathrm{p,0}$ appears here. For the latter we use a value of $\gamma_\mathrm{p,0} = 10^9~\mathrm{eV} / (m_\mathrm{p} c^2)$.\\
In this work we assume that cosmic rays consist only of protons and electrons. When the latter follow the same distribution as protons accelerated in supernova remnants one speaks of \textit{primary cosmic ray electrons}, which have a similar injection rate as the protons (\ref{Qp}). However, there is a second source of cosmic ray electrons: decay products of the cosmic ray protons. The latter can decay into neutral pions that further decay into electron positron pairs. Electrons, which have been produced in this way, are called \textit{secondary cosmic ray electrons}. The injection spectrum of secondaries can be estimated as \citep{LackiBeck2013}
\begin{equation}
  Q_\mathrm{e,sec} = \frac{f_\pi}{6} \frac{m_\mathrm{p}}{m_\mathrm{e}} \left(\frac{\gamma_\mathrm{p}}{\gamma_\mathrm{e}}\right)^2~Q_\mathrm{p}(\gamma_{\mathrm{p}}).
\label{Qesec}
\end{equation}
When assuming that the fraction of proton energy that goes into pion production $f_\pi = 0.2 - 0.5$ \citep{LackiEtAl2011}, the ratio of secondary to total cosmic ray electrons $Q_\mathrm{e}$, 
\begin{equation}
  f_\mathrm{sec} = \frac{Q_\mathrm{e,sec}}{Q_\mathrm{e}},
\label{fsec}
\end{equation}
is roughly 0.6 to 0.8. The energy of secondary electrons is $\gamma_\mathrm{e} = m_\mathrm{p}/(20 m_\mathrm{e}) \gamma_\mathrm{p}$ \citep{LackiBeck2013}.  \\
In our analysis the cosmic ray electrons play the most important role, as they are responsible for energy losses due to inverse Compton scattering. In order to find the steady-state spectrum of the electrons $N_\mathrm{e} (\epsilon_\mathrm{e})$ one needs to solve the diffusion-loss equation \citep{Longair2011}:
\begin{equation}
  \frac{\partial N(\gamma)}{\partial t} = Q(\gamma) + \frac{\mathrm{d}}{\mathrm{d} \gamma} \left[ b(\gamma) N(\gamma)\right] - \frac{N(\gamma)}{\tau(\gamma)} + D \nabla^2 N(\gamma).
\label{DiffLossEq}
\end{equation}
Here $Q(\gamma)$ is the injection spectrum, $b(\gamma) = - \mathrm{d} \gamma / \mathrm{d} t$ the energy loss rate, $\tau(\gamma$) the timescale of escape or total losses, and $D$ the spacial diffusion time scale. The diffusion loss equation is valid for electrons (index e) and for protons (index p). For a homogeneous medium, \citet{LackiBeck2013} find for the equilibrium proton spectrum, i.e.~$\partial N_\mathrm{p}(\gamma_\mathrm{p})/\partial t = 0$,
\begin{equation}
  N_\mathrm{p}(\gamma_\mathrm{p}) = Q_\mathrm{p}(\gamma_\mathrm{p})~f_\pi~\tau_\pi,
\label{Np}
\end{equation}
and for the electron spectrum, i.e.~$\partial N_\mathrm{e} (\gamma_\mathrm{e})/\partial t = 0$,
\begin{equation}
  N_\mathrm{e}(\gamma_\mathrm{e}) = \frac{Q_\mathrm{e}(\gamma_\mathrm{e})~\tau_\mathrm{e}(\gamma_\mathrm{e})}{\chi - 1}.
\label{Ne}
\end{equation}
The characteristic timescales appearing here are the one for pion production, 
\begin{equation}
  \tau_\pi = 50~\mathrm{Myr}~(n/\mathrm{cm}^{-3})^{-1},
\end{equation}
and the electron cooling time, $\tau_\mathrm{e} = \epsilon_\mathrm{e} / b_\mathrm{e}(\epsilon_\mathrm{e})$, which is calculated as
\begin{equation}
  \tau_\mathrm{e} = \left(\tau_\mathrm{synch}^{-1} + \tau_\mathrm{IC}^{-1} + \tau_\mathrm{ion}^{-1} + \tau_\mathrm{brems}^{-1}\right)^{-1}.
\label{te}
\end{equation}
This equation takes into account different energy losses of cosmic ray electrons, including synchrotron emission ($\tau_\mathrm{synch}$), inverse Compton scattering ($\tau_\mathrm{IC}$), ionization ($\tau_\mathrm{ion}$) and bremsstrahlung ($\tau_\mathrm{brems}$). We will discuss the importance of the different loss timescales in the next section. Combination of the upper equations yields for the steady-state spectra of cosmic ray protons and electrons
\begin{eqnarray}
  N_\mathrm{p}(\gamma_\mathrm{p}) & = & f_\pi \tau_\pi Q_{\mathrm{p},0} \gamma_\mathrm{p}^{-\chi} \label{Np0}\\
  N_\mathrm{e}(\gamma_\mathrm{e}) & = & \frac{20^{2-\chi} f_\pi}{6 f_\mathrm{sec} (\chi-1)} \tau_\mathrm{e} \left(\frac{m_\mathrm{e}}{m_\mathrm{p}}\right)^{1-\chi} Q_{\mathrm{p},0} \gamma_\mathrm{e}^{-\chi},
\label{Ne2}
\end{eqnarray}
with the normalization of the proton injection spectrum given in equation (\ref{Qp0}). \\
With the spectral distribution the energy density of cosmic rays can be computed. The dominant part of the total cosmic ray energy $U_\mathrm{CR}$ is contributed from the protons. It can be calculated by
\begin{equation}
  U_\mathrm{CR} = \int_{\gamma_\mathrm{p,0}}^\infty N_\mathrm{p}(\gamma_\mathrm{p})~\gamma_\mathrm{p} m_\mathrm{p} c^2~\mathrm{d} \gamma_\mathrm{p} + \int_{\gamma_\mathrm{e,0}}^\infty N_\mathrm{e}(\gamma_\mathrm{e})~\gamma_\mathrm{e} m_\mathrm{e} c^2~\mathrm{d} \gamma_\mathrm{e}.
\end{equation}
Note, that $U_\mathrm{CR}$ indicates the total energy, and not the energy density $u_\mathrm{CR}$. Assuming a Milky Way like galaxy with the properties given for a normal galaxy at $z=0$ in table \ref{Table_Props}, we find a total cosmic ray density of $u_\mathrm{CR} = 1.33~\mathrm{eV}\mathrm{cm}^{-3}$.



\section{Breakdown of the FIR-Radio Correlation at High Redshifts}
\label{Breakdown}

\begin{figure}
  \includegraphics[width=0.5\textwidth]{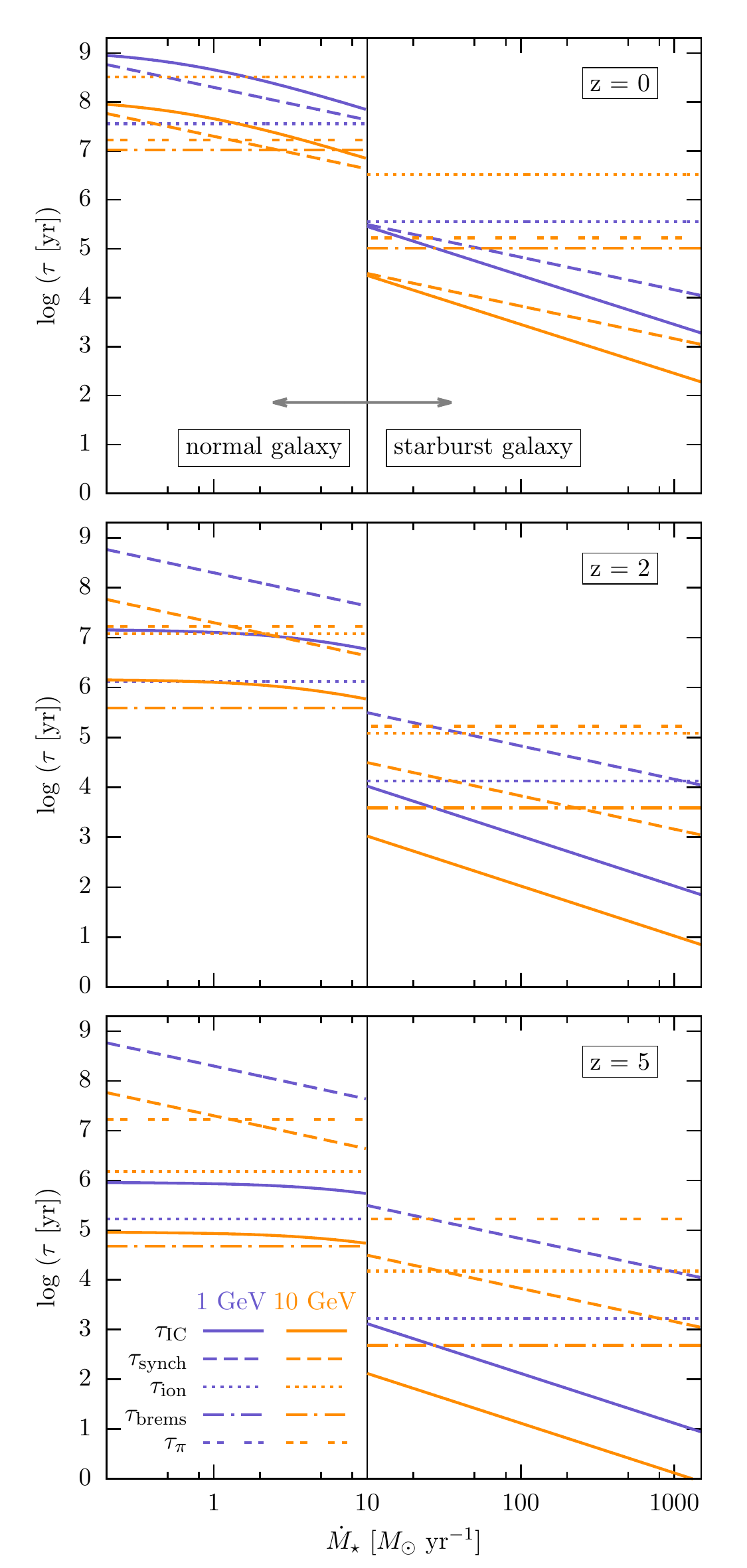}
  \caption{Typical timescales of cosmic ray electrons with an energy of 1 GeV (blue lines) and 10 GeV (orange lines) as a function of the star formation rate $\dot{M_\star}$. The different panels present the timescales at different redshifts $z$. The solid line indicates the timescale of inverse Compton scattering $\tau_\mathrm{IC}$, the dashed line synchrotron emission $\tau_\mathrm{synch}$, the dotted line ionization $\tau_\mathrm{IC}$, the dashed-dotted bremsstrahlung $\tau_\mathrm{brems}$ and the dashed-dotted pion production $\tau_\pi$. We show the case of a normal galaxy up to $10~\mathrm{M}_\odot \mathrm{yr}^{-1}$, for larger $\dot{M_\star}$ we use the starburst properties. The jump in timescales results from the different particle densities of the two galaxy models. Note, that $\tau_\mathrm{brems}$ and $\tau_\pi$ are independent of the cosmic ray energy and are thus superposed.}
\label{timescales_SFR}
\end{figure}

Observations of nearby galaxies show that there is a correlation between the far-infrared flux and the radio flux. \citet{YunEtAl2001} combine data from the NRAO VLA Sky Survey, which includes the 1.4 GHz radio luminosity, with the FIR luminosity data from the IRAS Redshift Survey. In their sample of around 1800 galaxies they find a tight correlation over roughly five orders of magnitude in luminosity.\\
The origin of the FIR-radio correlation can be understood as a result of coupling between star formation, cosmic rays and the magnetic field. The FIR radiation arises from the thermal emission of dust, which is heated by the ultraviolet radiation of massive stars, and thus traces the SFR. The SFR is also directly connected to the supernova rate, and hence to the cosmic ray production, which has been discussed before. The highly energetic cosmic rays lose their energy when traveling through the ISM. At present day one of the most important energy loss mechanism for cosmic ray electrons is synchrotron emission, which results from the interaction with interstellar magnetic field and lies in the radio regime.\\
Besides the synchrotron (synch) emission, the cosmic ray electrons can lose their energy also via inverse Compton scattering (IC), ionization (ion), and bremsstrahlung (brems), see also \citet{Oh2001}. The typical timescales of these processes are \citep{SchleicherBeck2013}:
\begin{eqnarray}
  \tau_\mathrm{synch} & = & \frac{3~m_\mathrm{e}~c}{4~\sigma_\mathrm{T}~u_\mathrm{B}~\gamma_\mathrm{e}} \\
  \tau_\mathrm{IC} & = & \frac{3~m_\mathrm{e}~c}{4~\sigma_\mathrm{T}~u_\mathrm{ISRF}~\gamma_\mathrm{e}} \\
  \tau_\mathrm{ion} & = & \frac{\gamma_\mathrm{e}}{2.7~c~\sigma_\mathrm{T}~(6.85 + 0.5~\mathrm{ln}\gamma_\mathrm{e})~n} \\
  \tau_\mathrm{brems} & = & 3.12 \times10^7~\mathrm{yr}~\left(\frac{n}{\mathrm{cm}^{-3}}\right)^{-1}.  
\end{eqnarray}
Here $\gamma_\mathrm{e}=\epsilon_\mathrm{e}/(m_\mathrm{e} c^2)$ is the Lorentz factor of a an electron with energy $\epsilon_\mathrm{e}$, $u_\mathrm{B} = B^2/(8 \pi)$ the magnetic energy density, $u_\mathrm{ISRF}$ the energy density of the ISRF (\ref{uISRFtot}), and $n$ the particle density of the interstellar medium (\ref{n}). \\
In figure \ref{timescales_SFR} we show the different timescales as a function of star formation rate at different redshifts. For calculating the timescale of synchrotron emission we use a scaling of the magnetic field strength of $B \propto \dot{M}_\star^{1/3}$, which has been observed in various galaxies \citep{NiklasBeck1997,ChyzyEtAl2011}. We use here $B_0=10^{-5}$ G for the normal galaxy and $B_0=2\times 10^{-4}$ G for the starburst galaxy, which is motivated from observations of M82 \citep{deCeadelPozoEtAl2009}. In the plot we present the timescales for two different electron energies, 1 GeV and 10 GeV, indicated by differently colored lines. Note, that the jump of the timescales in the transition from normal to starburst galaxies results from the increase of density in our model (see table \ref{Table_Props}). While losses by bremsstrahlung and synchrotron emission dominate in normal galaxies for low redshifts, in starburst galaxies inverse Compton scattering is the most important effect at high star formation rates already at $z=0$. With increasing redshift figure \ref{timescales_SFR} illustrates that the difference between synchrotron and inverse Compton increases continuously.  \\
Form the analysis of the timescales we conclude that inverse Compton scattering is the dominant loss mechanism of cosmic ray electrons for normal galaxies only at high redshifts ($z \gtrsim 5$) and in starburst galaxies with high SFRs at basically all cosmological times. The typical frequency of inverse Compton scattered photons with the CMB lies in the X-ray regime. Thus, we expect galaxies to become bright in the X-ray above a critical redshift and star formation rate. In the next section we present a model for the typical X-ray emissivity due to inverse Compton scattering.



\section{Expected X-Ray Luminosity from Inverse Compton Scattering}  
\label{ExpectedIC}

\subsection{Characteristic Frequencies}
\label{ExpectedIC_CharacFreq}

In the process of inverse Compton scattering a low energy photon scatters on an high energy electron. Due to the radiation field the electron gets decelerated and transmits energy on the photon. The characteristic frequency of an inverse Compton scatted photon, which we observe today, is
\begin{equation}
  \nu_\mathrm{charac}(z) = \gamma_\mathrm{e,0}^2 \frac{\nu_\mathrm{in}}{1+z},
\label{ICcharac}
\end{equation}
where $\gamma_\mathrm{e,0}$ is the typical energy of cosmic ray electrons and $\nu_\mathrm{in}$ is the frequency of the incoming photon. \\
The energy range of the \textit{Chandra} telescope, which we will use in our calculation if not otherwise indicated, is
\begin{eqnarray}
  \nu_\mathrm{C1} & = & 0.5 ~\mathrm{keV} \\
  \nu_\mathrm{C2} & = & 8 ~\mathrm{keV}. 
\end{eqnarray}
Thus, photons with frequencies between
\begin{equation}
  \nu_\mathrm{in,1} = \gamma_\mathrm{e,0}^{-2}~\nu_\mathrm{C1}~(1+z) \approx 3.16\times10^{10}~(1+z)~\mathrm{Hz}
\end{equation}
and 
\begin{equation}
  \nu_\mathrm{in,2} = \gamma_\mathrm{e,0}^{-2}~\nu_\mathrm{C2}~(1+z) \approx 6.31\times10^{11}~(1+z)~\mathrm{Hz}
\end{equation}
can be inverse Compton scattered into the detection range. We indicate the range of suitable frequencies in figures \ref{ISRF_nu} and \ref{Efficiency_nuin} by vertical lines. The most important components are thus the CMB and the IR component. As there is a distribution in the incoming photon energy as well as in the the cosmic ray energy, also photons with an initially different frequency can be scattered into the \textit{Chandra} range. However, we expect the majority of detected photons to origin from the calculated energy regime.

\subsection{Inverse Compton Luminosity}

In a galaxy, the initial photon spectrum can be approximated as the sum of several blackbody spectra, which we have modeled in (\ref{nISRF}), and the initial electron spectrum is a power-law (\ref{Qesec}). \citet{BlumenthalGould1970} show that the spectral distribution of inverse Compton scattered photons, i.e.~the total number of photons that are scattered into the energy $\epsilon = h \nu$, is 
\begin{eqnarray}
  Q_{\mathrm{IC}}(\nu) = \int_{0}^\infty \int_{\gamma_\mathrm{min}}^\infty N_{\mathrm{e}}(\gamma_\mathrm{e}) Q_{\mathrm{IC,e}}(\gamma_\mathrm{e}, \nu_\mathrm{in})~\mathrm{d}\gamma_\mathrm{e}~\mathrm{d}\nu_\mathrm{in}, 
\label{Qint1}
\end{eqnarray}
with the contribution of a single electron of energy $\epsilon_\mathrm{e}=\gamma_\mathrm{e} m_\mathrm{e} c^2$ being
\begin{eqnarray}
  Q_{\mathrm{IC,e}}(\gamma_\mathrm{e}, \nu_\mathrm{in}) & = & \frac{\pi r_0^2 c h}{2 \gamma_\mathrm{e}^4} \frac{n_{\mathrm{ISRF},\nu}(\nu_\mathrm{in})}{\nu_\mathrm{in}^2} \left( 2 \nu \mathrm{ln}\left(\frac{\nu}{4 \gamma_\mathrm{e}^2 \nu_\mathrm{in}}\right) \right.  \nonumber \\
                                                        &   & \left. + \nu + 4 \gamma_\mathrm{e}^2 \nu_\mathrm{in} - \frac{\nu^2}{2 \gamma_\mathrm{e}^2 \nu_\mathrm{in}}\right). 
\label{Qintsing}
\end{eqnarray}
This result is valid in the so-called Thomson limit, where the energy of the photon before scattering $h\nu \approx k T(z)$ is much less then the rest energy of the electron $m_e c^2$. While $m_e c^2$ is roughly $10^{-6}$ erg, $k T(z)$ varies from a few times $10^{-16}$ erg to a few times $10^{-15}$ erg in the redshift range $z=0-10$. Thus, the Thomson limit is valid within our model. \\
In the Thomson limit the lower integration limit in (\ref{Qint1}) is $\gamma_\mathrm{min} = 1/2~(\nu/\nu_\mathrm{in})^{1/2}$. Here $\nu_\mathrm{in}$ is the frequency of the incoming photon, while $\nu$ is the frequency of the inverse Compton scattered photon. With the spectrum of cosmic ray electrons (\ref{Ne}) and the interstellar radiation field (\ref{nISRF}), integration over $\gamma_\mathrm{e}$ yields again to a power-law in the frequency
\begin{eqnarray}
  Q_{\mathrm{IC}}(\nu) & = & \frac{\pi h r_0^2 c^4 m_\mathrm{e}^{2-\chi} m_\mathrm{p}^\chi f_\pi}{f_\mathrm{sec} \sigma_\mathrm{T}  u_\mathrm{ISRF}} Q_{\mathrm{p},0}~\tilde{F}(\chi)   \nonumber \\
                       &   & \times (h \nu)^{-(\chi+2)/2}   \nonumber \\
                       &   & \times \int_0^\infty (h \nu_\mathrm{in})^{\chi/2} n_{\mathrm{ISRF},\nu}(\nu_\mathrm{in})~\mathrm{d}\nu_\mathrm{in}, 
\label{Qint}
\end{eqnarray}
with the abbreviation
\begin{eqnarray}
  \tilde{F}(\chi) = \frac{(\chi^2+6\chi+16) 2^{5-\chi} 5^{2-\chi}}{(4+\chi)^2 (\chi+6) (\chi+2) (\chi-1)}. 
\end{eqnarray}
Note, that here the normalization $Q_\mathrm{p,0}$ (\ref{Qp0}) of the injection spectrum of cosmic ray protons enters.

\begin{figure}
  \includegraphics[width=0.5\textwidth]{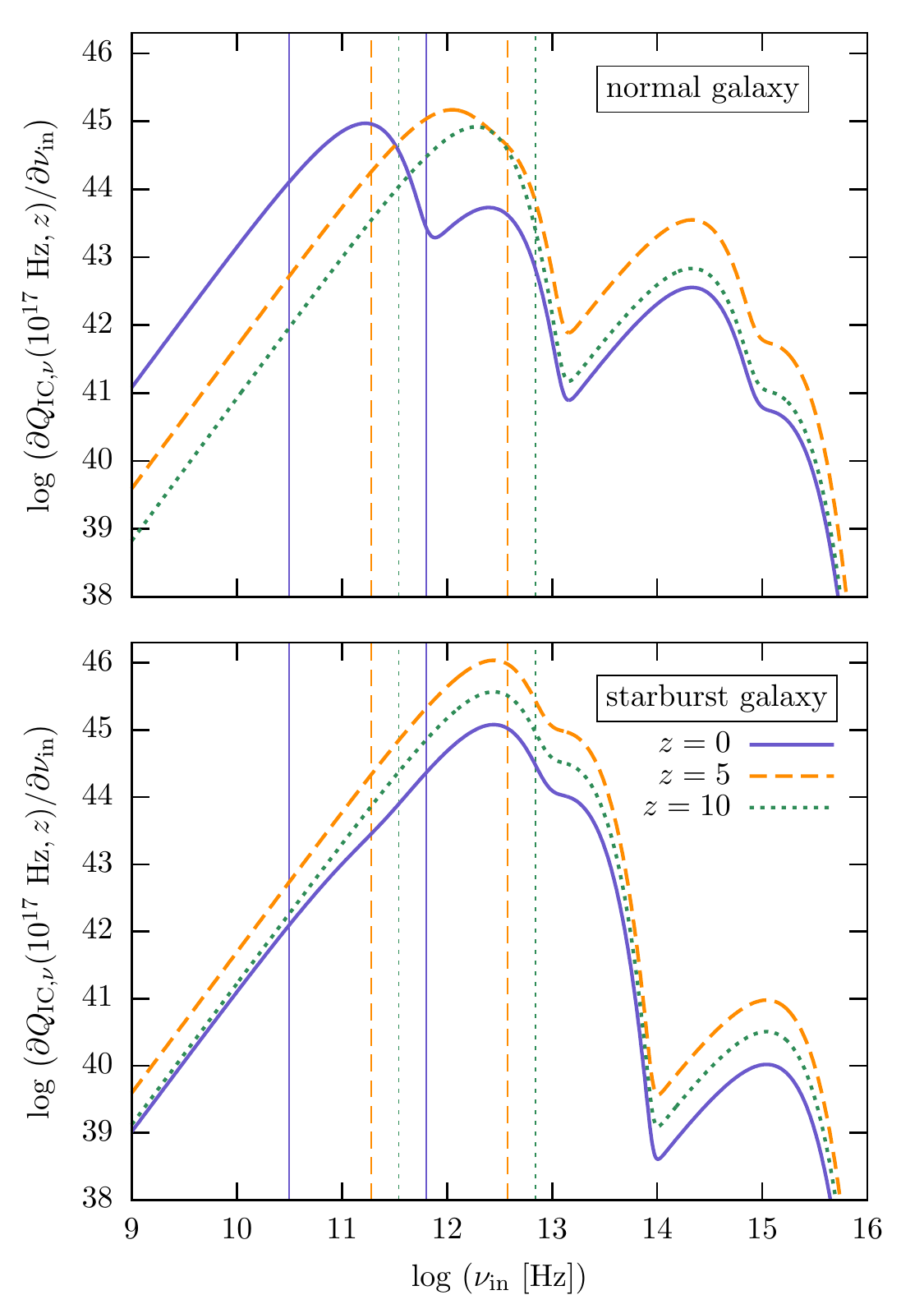}
  \caption{The efficiency of inverse Compton scattering for our model of the interstellar radiation field as a function of the incoming photon frequency $\nu_\mathrm{in}$. The different line styles indicate the efficiency at different redshifts: The solid blue lines give $z=0$, the dashed orange lines $z=5$ and the dotted green lines $z=10$. The upper panel shows the result for a galaxy with a normal SFR, the lower panel the one for a starburst galaxy. We use here the redshift-dependent SFR model (\ref{SFR}) and factor of $\xi = 0.1$ in the normalization of the cosmic ray spectrum (\ref{Qp0}). The different components of the ISRF are listed in table \ref{Table_ISRF}. The vertical lines indicate typical frequency range of photons that are inverse Compton scattered into the X-ray regime.\vspace{0.3cm}}
\label{Efficiency_nuin}
\end{figure}

\begin{figure}
  \includegraphics[width=0.5\textwidth]{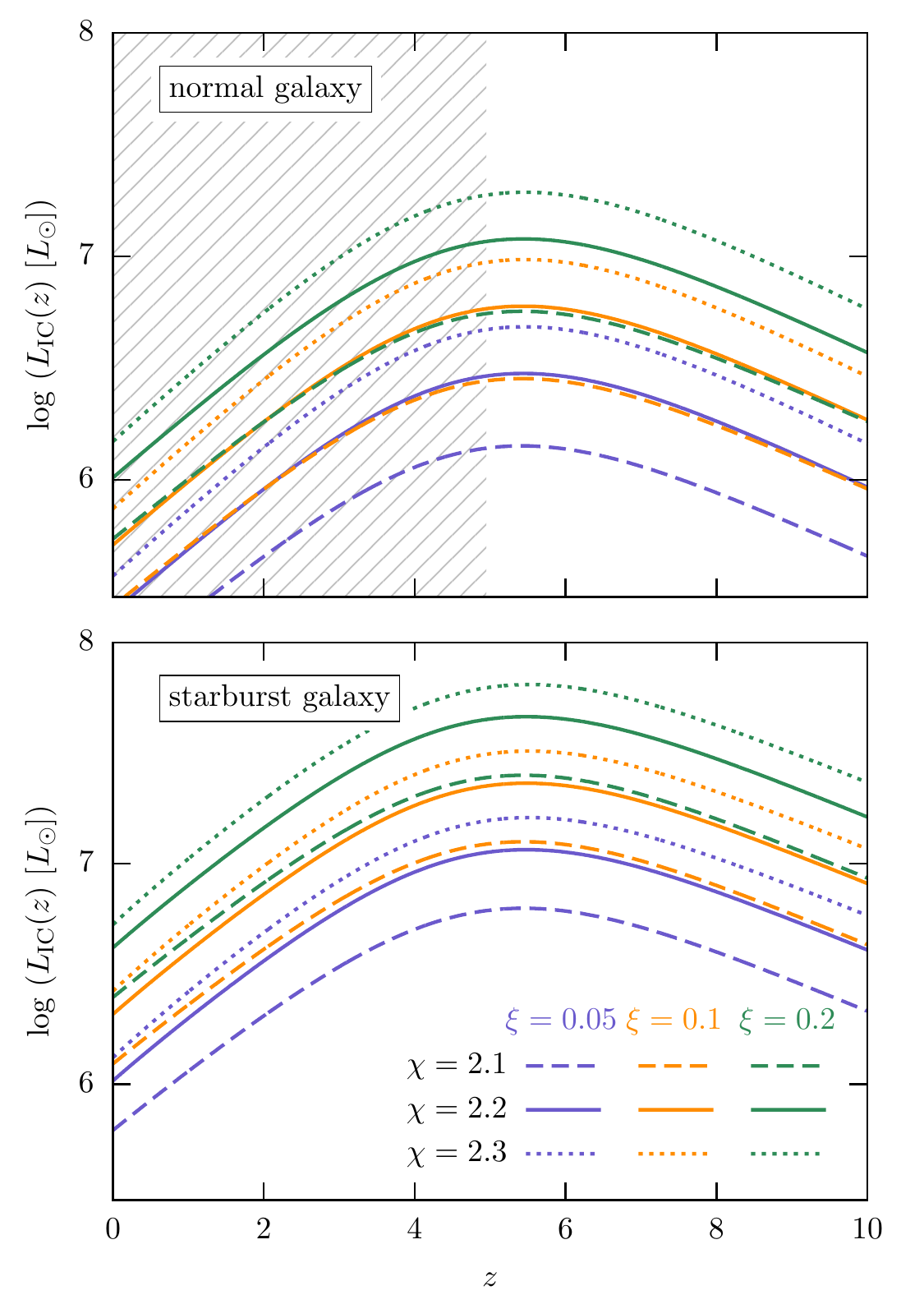}
  \caption{X-ray Luminosity $L_\mathrm{IC}$ resulting from inverse Compton scattering as a function of redshift $z$. We use the redshift dependent SFR model (\ref{SFR}), and show the case of a normal galaxy in the upper panel and the one for a starburst galaxy in the lower panel. The influence of different cosmic ray spectra (see equation \ref{Ne}) is tested. The different line colors represent calculations with different normalizations $\xi$ ranging from $\xi = 0.05$ (blue lines) over $\xi = 0.1$ (orange lines) up to $\xi = 0.2$ (green lines). In each case we also change the slope of the cosmic ray spectrum $\chi$. The dashed lines are results for $\chi = 2.1$, the solid lines for $\chi =2.2$ and the dotted lines $\chi = 2.3$. Note, that we do not show the luminosity for a normal galaxy up to a redshift of 5, as here inverse Compton scattering is not the dominant loss mechanism of cosmic ray electrons. \vspace{0.3cm}}
\label{L_z__z}
\end{figure}

From equation (\ref{Qint}) we can calculate the spectral contribution to the inverse Compton scattering $\partial Q_{\mathrm{IC},\nu}(\nu,z)/\partial \nu_\mathrm{in}$, which is shown in figure \ref{Efficiency_nuin}. For a typical resulting frequency we chose $\nu=10^{17}$ Hz, which is motivated by equation (\ref{ICcharac}) when using an incoming frequency of $\nu_\mathrm{in}=10^{11}$ Hz and an electron energy of $\gamma_\mathrm{e} = 10^{9} \mathrm{eV} / (m_\mathrm{e} c^2)$ Hz. In the top panel of figure \ref{Efficiency_nuin} we show the efficiency for the ISRF of a normal galaxy as a function of frequency. In the case of $z=0$ one can clearly distinguish the contributions of the different ISRF components. From the left to the right one can see the CMB component, the (cold) IR component, the optical component and the UV component, while the CMB component dominates clearly. If we go to higher redshifts the temperature and with that the peak of the CMB shifts to larger frequencies. The contribution of the CMB becomes more and more important. Already at $z=5$ the IR component is barley visible anymore. To conclude, in a galaxy with normal star formation the CMB is the dominant incoming radiation field for inverse Compton scattering. For comparison we show the inverse Compton efficiency of a starburst galaxy in the lower panel of figure \ref{Efficiency_nuin}. Here the cold IR component is the dominant one for the inverse Compton mechanism. Even for the $z=0$ case, the peak of the CMB is almost not visible. Due to the strong IR radiation field we expect starburst galaxies to be more efficient in inverse Compton scattering.\\
Finally, we are interested in the luminosity that results from inverse Compton scattering. For this we continue with the evaluation of the integral in equation (\ref{Qint}), which yields
\begin{eqnarray}
  Q_{\mathrm{IC},\nu}(\nu,z) & = & \frac{4 \pi r_0^2 c k^3 m_\mathrm{e}^{2-\chi} m_\mathrm{p}^\chi f_\pi}{f_\mathrm{sec} h^2 \sigma_\mathrm{T}  u_\mathrm{ISRF}}~F(\chi)~Q_{\mathrm{p},0}  \nonumber \\
                             &   & \times (h \nu)^{-(2+\chi)/2} ~\sum_i f_i~\left(k T_i\right)^{\chi/2}.
\label{Qnu}
\end{eqnarray}
Here we introduced the abbreviation
\begin{eqnarray}
  F(\chi)  & = & \Gamma\left(\frac{6+\chi}{2}\right)  \zeta \left(\frac{6+\chi}{2}\right)~\tilde{F}(\chi).
\label{Fchi}
\end{eqnarray} 
With the spectral distribution of inverse Compton scattered photons (\ref{Qnu}) we can calculate the spectral luminosity
\begin{equation}
  L_{\mathrm{IC},\nu}(\nu,z) = Q_{\mathrm{IC},\nu}(\nu,z)~ h \nu
\end{equation}
and the integrated X-ray luminosity
\begin{eqnarray}
  L_\mathrm{IC}(z) & = & \int_{\nu_1}^{\nu_2} L_{\mathrm{IC},\nu}(\nu,z)~\mathrm{d}\nu \nonumber \\
                   & = & \frac{4 \pi r_0^2 c k^3 m_\mathrm{e}^{2-\chi} m_\mathrm{p}^\chi f_\pi}{f_\mathrm{sec} h^{(4+\chi)/4} \sigma_\mathrm{T}  u_\mathrm{ISRF}}~\frac{2 F(\chi)}{2-\chi}~Q_{\mathrm{p},0} \nonumber \\
                   &   & \times \left(\nu_2^{1-\chi/2}-\nu_1^{1-\chi/2}\right)~\sum_i f_i~\left(k T_i\right)^{\chi/2}.   
\label{L_IC}
\end{eqnarray}
The integrated X-ray luminosity that purely results from the inverse Compton scattering as a function of redshift $z$ is shown in figure \ref{L_z__z}. For this plot we use the redshift dependent model of the ISRF given in table \ref{Table_ISRF} with the SFR model from equation (\ref{SFR}). We test here the influence of the cosmic ray spectrum, which is given in equation (\ref{Ne}). The normalization of the cosmic ray spectrum $\xi$ is varied over one order of magnitude from $\xi = 0.05$ to $0.2$. From the figure one notes that with increasing $\xi$ the luminosity increases, which can also be seen from equation (\ref{Qnu}) directly, where the total number of injected cosmic ray protons $Q_{\mathrm{p},0}$ appears, which is proportional to $\xi$ (see equation \ref{Np0}). It is intuitively clear that with a larger number of cosmic ray protons the number of cosmic ray electrons and thus the number of scattering events increases, which leads to a larger inverse Compton luminosity. In figure \ref{L_z__z} we also test the influence of changing the slope of the cosmic ray spectrum $\chi$, which we vary from 2.1 to 2.3. We expect that this should not change significantly as the basic cosmic ray acceleration mechanism should be the same for all galaxies and redshifts. With an increasing $\chi$ the luminosity decreases. \\
We show in figure \ref{L_z__z} the case of a normal galaxy in the upper panel. Here we delete the range from $z=0$ to $z=5$, as in this range inverse Compton scattering is not the dominant energy loss of cosmic ray electrons and thus our model can not be applied in the present form (see discussion in section \ref{Breakdown}). The lower panel shows the inverse Compton luminosity of a starburst galaxy. In both cases the evolution of the luminosity follows closely the one of the star formation rate, which explains the peak at $z\approx 5$ as well as the larger luminosity of the starburst galaxy.

\subsection{Inverse Compton Flux}

\begin{figure}
  \includegraphics[width=0.5\textwidth]{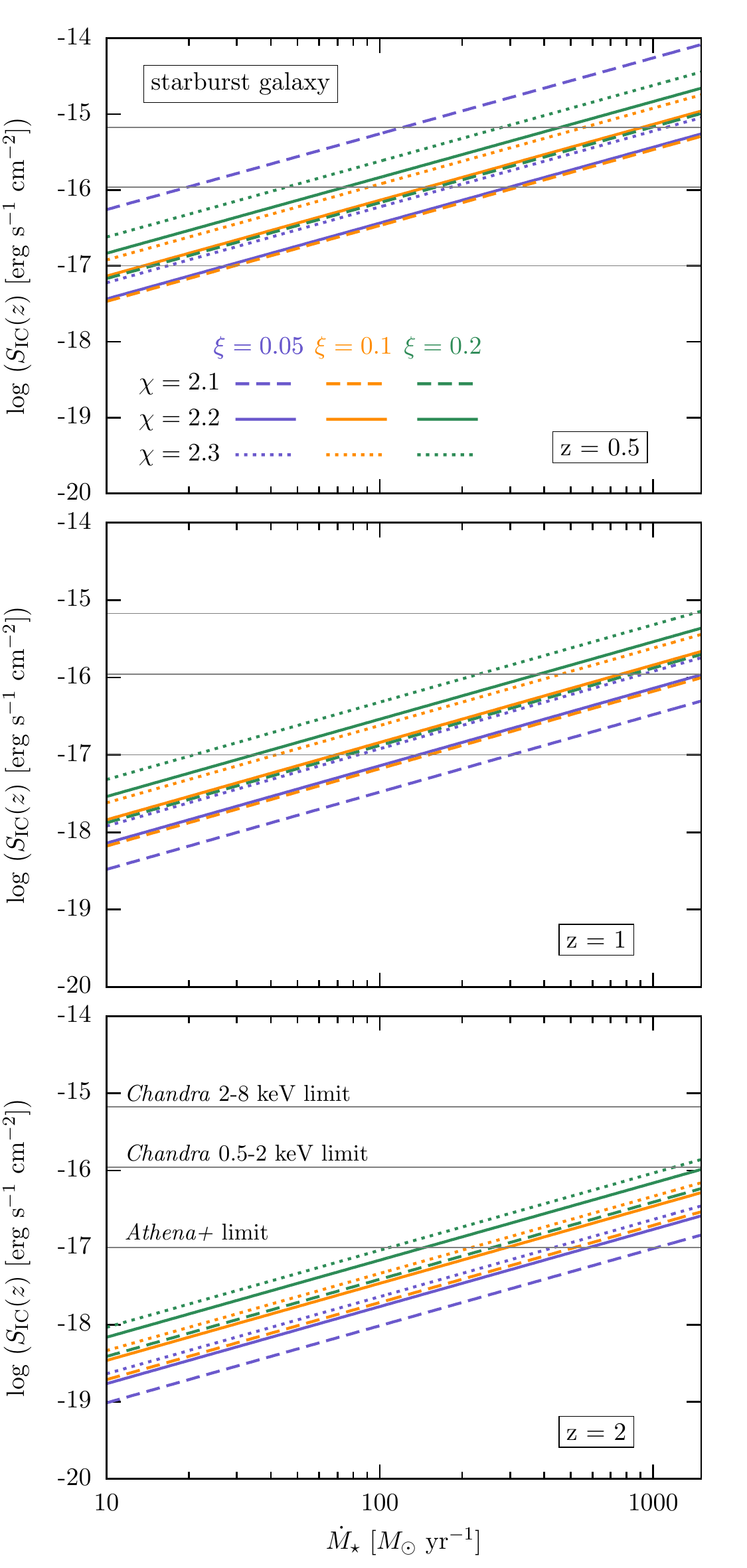}
  \caption{The observed flux of inverse Compton scattering in the range 0.5 to 8 keV from a starburst galaxy as a function of star formation rate $\dot{M}_\star$. The different panels represent different redshifts from $z =0.5$ to $z=2$. The line colors indicate the normalization of the cosmic ray spectrum $\xi$ (see equation \ref{Np0}), while the linestyles show different slopes of the cosmic ray injection spectrum: $\chi = 2.1$ (dashed lines), $\chi = 2.2$ (solid lines) and $\chi = 2.3$ (dotted lines). We further indicate the sensitivity limits of \textit{Chandra} and \textit{Athena+} by the horizontal lines. \vspace{0.3cm}}
\label{SIC_SFR}
\end{figure}
With the luminosity distance,
\begin{equation}
  d_\mathrm{L}(z) = (1+z)^2~d_\mathrm{A}(z),
\end{equation}
where the angular diameter distance $d_\mathrm{A}$ can be determined from the Mattig relation,
\begin{eqnarray}
  d_\mathrm{A}(z) & = & \frac{c}{H_0} \frac{2}{\Omega_\mathrm{m}^2 (1+z)^2} \left(\Omega_\mathrm{m} z + (\Omega_\mathrm{m}-2) \right. \nonumber \\
   &   &  \left. \times (\sqrt{1+\Omega_\mathrm{m}z}-1)\right),
\end{eqnarray}
we can moreover calculate the flux resulting from inverse Compton scattering. We use the latest cosmological parameters determined by the \textit{Planck} satellite. The Hubble constant is $H_0 = 67.11~\mathrm{km~s}^{-1} \mathrm{Mpc}^{-1}$ and the matter density parameter $\Omega_\mathrm{m} = 0.3175$ \citep{PlanckCollaboration2013}. We determine the spectral flux density by
\begin{equation}
  S_{\mathrm{IC},\nu} = \frac{L_{\mathrm{IC},\nu}}{4 \pi d_\mathrm{L}^2}
\label{SICnu}
\end{equation}
and the total flux density by
\begin{equation}
  S_{\mathrm{IC}}(z) = \int_{\nu_1}^{\nu_2} S_\nu(\nu,z)~\mathrm{d}\nu.
\end{equation}
We show the resulting inverse Compton flux from a starburst galaxy in figure \ref{SIC_SFR} as a function of a fixed SFR. The different line styles and colors cover our parameter space of the cosmic ray spectrum. We again vary $\xi$ from 0.05 to 0.2 and $\chi$ from 2.1 to 2.3. The horizontal lines in the plot give the sensitivity limits of \textit{Chandra} and the future X-ray observatory \textit{Athena+}. For the Extended \textit{Chandra} Deep Field the flux limit in the 0.5-2 keV range is $1.1\times10^{-16}~\mathrm{erg~s}^{-1}\mathrm{cm}^{-3}$ and $6.7\times10^{-16}~\mathrm{erg~s}^{-1}\mathrm{cm}^{-3}$ in the 2-8 keV range \cite{LehmerEtAl2005}. The expected sensitivity limit for \textit{Athena+} is $10^{-17}~\mathrm{erg~s}^{-1}\mathrm{cm}^{-3}$ in the 0.5-2 keV band \citep{NandraEtAl2013}. With increasing redshift the inverse Compton flux moves more and more out of the detection limits. However, with \textit{Athena+} the pure inverse Compton flux of objects with high SFRs should still be visible at redshifts larger than 2.



\section{Distinguishing Other X-Ray Processes}
\label{OtherXrays}

A typical galaxy contains various sources of X-ray emission \citep{PersicRephaeli2002}. While normal stars only contribute a small fraction to the total X-ray emission, supernova remnants and the hot thermal ISM gas are more important. It has been shown, however, that the dominant sources are X-ray binaries. In the following we present two methods from literature, which estimate the X-ray emission from X-ray binaries. We further shortly discuss the X-ray emission from supernovae and active galactic nuclei.

\subsection{X-Ray Binaries}

\subsubsection{Analytical Model for the Mean Evolution of X-Ray Binaries}
\label{AnaXBs}
\begin{figure}
  \includegraphics[width=0.5\textwidth]{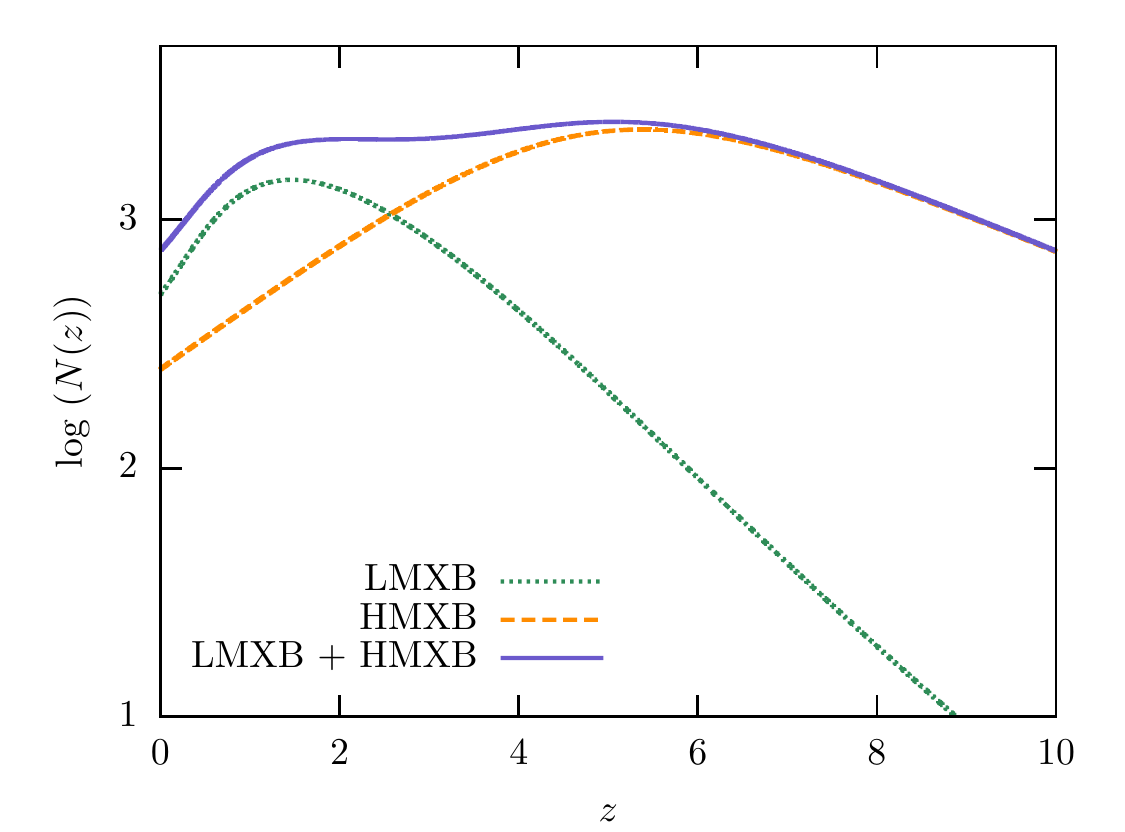}
  \caption{The number of X-ray binaries in a starburst galaxy as a function of redshift $z$. We show the number of low-mass X-ray binaries (LMXB, orange lines), high-mass X-ray binaries (HMXB, green lines) and the sum of both (LMXB+HMXB, blue lines). \vspace{0.3cm}}
\label{NXB_z}
\end{figure}

Observations show that the emission of a normal galaxy is dominated by a few point sources, which have been identified as X-ray binaries \citep{Fabbiano1995}. Any model of the X-ray emission of a normal galaxy should thus show a characteristic scaling with the number of X-ray binaries and also with the SFR.\\
\citet{GhoshWhite2001} provide an analytical model for the number of the different X-ray binary classes. The evolution of the number of high-mass X-ray binaries (HMXB) in a typical galaxy $N_\mathrm{HMXB}$ is governed by
\begin{equation}
  \frac{\partial N_\mathrm{HMXB}(t)}{\partial t} = \alpha_\mathrm{HMXB} \frac{\dot{M}_{\star}}{10 \mathrm{M}_\odot} - \frac{N_\mathrm{HMXB}(t)}{\tau_\mathrm{HMXB}},
\end{equation}
where the typical HMXB evolution timescale $\tau_\mathrm{HMXB}$ is $5\times10^6$ yr. The parameter $\alpha_\mathrm{HMXB}$ gives the rate of HMXB formation and can be estimated by $\alpha_\mathrm{HMXB}\approx \frac{1}{2} f_\mathrm{binary} f_\mathrm{prim}^\mathrm{HMXB} f_\mathrm{SN}^\mathrm{HMXB}$. Here, $f_\mathrm{binary}$ is the fraction of stars that are in binaries, $f_\mathrm{prim}$ is the fraction of binaries, that are in the right mass range for evolving into a X-ray binary and $f_\mathrm{SN}$ is the fraction of the binary systems that survive the first supernova explosion. \\
The evolution of low-mass X-ray binaries (LMXB) is more complicated due to the similarity of the post supernova binary (PSNB) and the real LMXB timescales, $\tau_\mathrm{PSNB}\approx 1.9\times10^9$ yr and $\tau_\mathrm{LMXB}\approx 10^9$ yr. The evolution of the total abundances $N_\mathrm{PSNB}$ and $N_\mathrm{LMXB}$ is described by the following coupled equations:
\begin{eqnarray}
  \frac{\partial N_\mathrm{PSNB}(t)}{\partial t} & = & \alpha_\mathrm{PSNB} \frac{\dot{M}_{\star}}{10 \mathrm{M}_\odot} - \frac{N_\mathrm{PSNB}(t)}{\tau_\mathrm{PSNB}} \\
  \frac{\partial N_\mathrm{LMXB}(t)}{\partial t} & = & \frac{N_\mathrm{PSNB}(t)}{\tau_\mathrm{PSNB}} - \frac{N_\mathrm{LMXB}(t)}{\tau_\mathrm{LMXB}}.
\end{eqnarray}
Here the parameter $\alpha_\mathrm{PSNB}$ is defined as $\frac{1}{2} f_\mathrm{binary} f_\mathrm{prim}^\mathrm{LMXB} f_\mathrm{SN}^\mathrm{LMXB}$. \\
The values of the different fractions $f$ (especially as a function of redshift) are very hard to estimate \citep{Fabbiano1995}. We thus use the observed number of X-ray binaries to calibrate $\alpha_\mathrm{HMXB}$ and $\alpha_\mathrm{LMXB}$ at $z=0$. With a value of roughly 100 LMXBs and 50 HMXB in the Milky Way \citep{GrimmEtAl2002} we find $\alpha_\mathrm{HMXB}\approx 5.00 \times 10^{-5}$ and $\alpha_\mathrm{LMXB}\approx 4.35 \times 10^{-7}$. The resulting numbers of HMXBs and LMXBs are shown in figure \ref{NXB_z}. We present here the evolution for the case of a starburst galaxy, which is expected to have more X-ray binaries due to a larger SFR. Note, that the evolution of the X-ray binary population follows closely the history of star formation. The number of HMXBs peaks at a redshift of roughly 5, which corresponds to the peak of the star formation rate (see figure \ref{VgalSFR_z}), while the peak of the LMXB is at a smaller redshift of roughly 1.5. This evolutionary delay comes from the long LMXB timescales, $\tau_\mathrm{PSNB}$ and $\tau_\mathrm{LMXB}$, and results in the fact that the LMXBs are the dominant type of X-ray binaries at present day. \\
For computing the X-ray luminosity that results from the X-ray binaries in a galaxy, we need to know the typical luminosities of LMXBs and HMXBs. In a detailed study of the Milky Way \citet{GrimmEtAl2002} find that the total luminosity of all X-ray binaries in the 2-10 keV range is $\approx 2-3 \times 10^{39}~\mathrm{erg~s}^{-1}$ (LMXB) and $\approx 2-3 \times 10^{38}~\mathrm{erg~s}^{-1}$ (HMXB). With the total numbers of the X-ray binaries given above this corresponds to mean luminosities of $\approx 2.5 \times 10^{37}~\mathrm{erg~s}^{-1}$ (LMXB) and $\approx 5.0 \times 10^{36}~\mathrm{erg~s}^{-1}$ (HMXB) in the 2-10 keV band. In our study we use the 0.5-8 keV band and thus need to estimate the luminosity in this band. \citet{WangEtAl2013} find a typical conversion factor of 1.21 between the two bands for X-ray binaries, which we apply here, too. We thus find for the total luminosity due to X-ray binaries in the 0.5-8 keV band
\begin{eqnarray}
  L_{XB}(z) & = & 1.21\times2.5\times10^{37}~\mathrm{erg~s}^{-1}~N_\mathrm{LMXB}(z) \nonumber \\
            &   & +~1.21\times5.0\times10^{36}~\mathrm{erg~s}^{-1}~N_\mathrm{HMXB}(z).
\end{eqnarray}
The evolution of $L_{XB}$ according the analytical model is presented in figure \ref{LXB_z}, where we also plot the luminosities from inverse Compton scattering of a starburst for comparison. 

\subsubsection{Observational Relations of X-Ray Binaries}
\label{ObsXBs}
As HMXBs evolve on very short timescales (see above), they are a good tracer for the SFR. However, at SFRs comparable to the Milky Way or lower, the X-ray binary population is dominated by LMXBs, which in return provide a measure for the total stellar mass in a galaxy. Only in galaxies with high SFRs the X-ray luminosity from X-ray binaries is dominated by the emission of HMXBs \citep{PersicRephaeli2007}. \\
For the X-ray binary emission, \citet{LehmerEtAl2010} find the following correlation with the star formation rate
\begin{eqnarray}
  L_{\mathrm{XB}}^{2-10~\mathrm{keV}} = 10^{39.43}~\mathrm{erg~s}^{-1}~\left(\frac{\dot{M}_\star}{\mathrm{M}_\odot \mathrm{yr}^{-1}}\right)^{0.74}.
\end{eqnarray}
This correlation is valid for the $2-10~\mathrm{keV}$ range and Kroupa IMF. Converting into the $0.5-8~\mathrm{keV}$ the X-ray luminosity changes to \citep{WangEtAl2013}
\begin{eqnarray}
  L_\mathrm{XB}^{0.5-8~\mathrm{keV}} & = & 1.21\times10^{39.43}~\mathrm{erg~s}^{-1}~\left(\frac{\dot{M}_\star}{\mathrm{M}_\odot \mathrm{yr}^{-1}}\right)^{0.74}. \nonumber \\
 & &
\label{LXBSFR}
\end{eqnarray}


\subsection{Supernova Remnants}

\begin{figure}
  \includegraphics[width=0.5\textwidth]{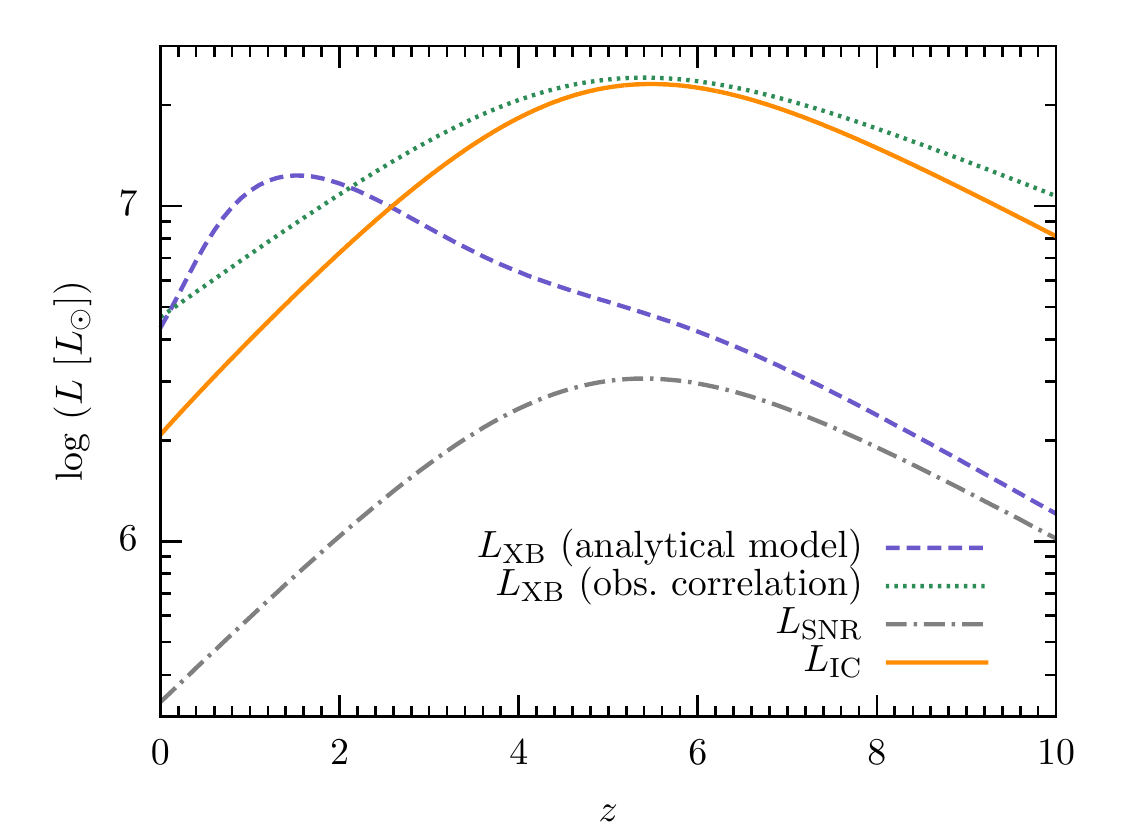}
  \caption{The luminosities of various X-ray sources as a function of redshift $z$. The luminosity from X-ray binaries is shown by the dashed blue line, when using the analytical evolutionary model, and by the dotted green line, when using the observational correlation. The X-ray luminosity contribution from supernova remnants are represented by the dashed-dotted gray line. For comparison we show the inverse Compton scattering luminosity (solid orange line) for our fiducial cosmic ray spectrum ($\xi = 0.1$ and $\chi=2.2$). We use here the redshift dependent ISRF from table \ref{Table_ISRF} and a starburst galaxy. \vspace{0.3cm}}
\label{LXB_z}
\end{figure}

Besides X-ray binaries, supernova remnants are point sources of X-ray emission in galaxies. The thermal X-ray radiation is emitted mostly during the free expansion and the Sedov-Taylor phase with a typical duration of less then $\tau_\mathrm{SNR} = 10^3$ yr \citep{Woltjer1972,Chevalier1977}. The typical number of X-ray emitting supernova remnants that are observed in a galaxy is
\begin{eqnarray}
  N_\mathrm{SNR} = \tau_\mathrm{SNR} \dot{N}_\mathrm{SN} = 12.72~\frac{\dot{M}_\star}{\mathrm{M}_\odot \mathrm{yr}^{-1}},
\end{eqnarray}
where we use the supernova rate (\ref{SNrate}). With a typical X-ray luminosity of supernova remnants of $10^{37}~\mathrm{erg~s}^{-1}$ the total X-ray emission of galaxies with star formation rates between 10 and 1500 $\mathrm{M}_\odot \mathrm{yr}^{-1}$ are roughly $10^4 - 10^6~\mathrm{L}_\odot$. Compared to the expected luminosity of X-ray binaries presented for example in figure \ref{LXB_z}, supernova remnants provide only a minor contribution to a galaxy's total X-ray emission.



\subsection{Active Galactic Nuclei}

If a galaxy hosts an active galactic nucleus (AGN) we expect additional X-ray emission. Typical X-ray luminosities of nearby AGNs are around $10^7~\mathrm{L}_\odot$. We expect however, that a large fraction of the radiation from the central black hole is absorbed by dust. We thus use the observed values of the X-ray luminosity that are not corrected for dust absorption. Ideal for our analysis would be starburst galaxies at high redshifts without AGNs, which might be detected in future observations.



\section{Application to Exemplary Galaxies}
\label{Examples}

\begin{figure}
  \includegraphics[width=0.5\textwidth]{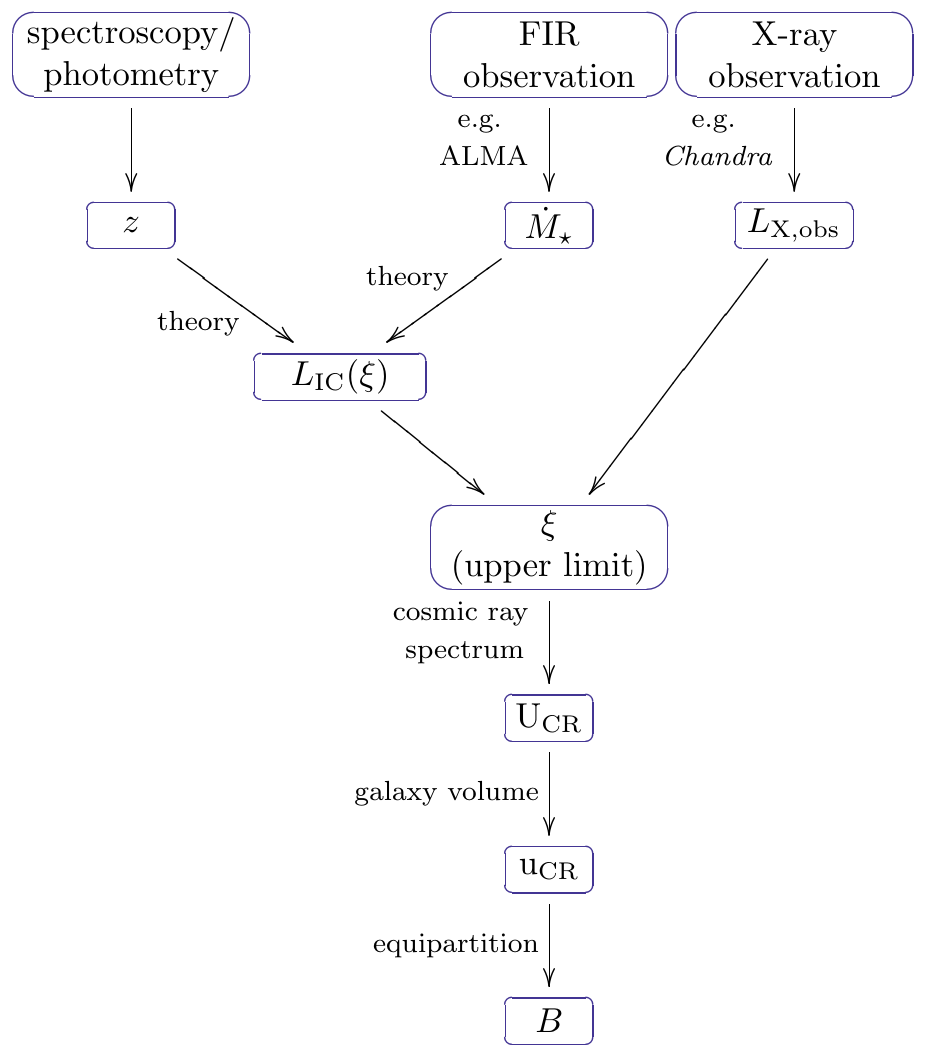}
  \caption{Illustration of our strategy for determining the cosmic ray density the magnetic field strength in redshifted star-forming galaxies from observations of the redshift, the far-infrared and the X-ray flux. For more details see section \ref{Examples}.\vspace{0.2cm}}
\label{Strategy}
\end{figure}

In this section we apply our model for the inverse Compton scattering to real galaxies in order to determine properties of cosmic rays and magnetic fields. Our strategy is illustrated in figure \ref{Strategy}. As observational input we need the redshift $z$ of an object, the FIR luminosity and the X-ray luminosity. The redshift can be determined spectroscopically or with photometry. For the FIR luminosity of objects at high redshift, from which we can determine the SFR $\dot{M}_\star$, there is a lot of data available from several surveys and also \textit{ALMA} will be a powerful tool in the future, and we can use \textit{Chandra} data for the X-ray luminosity $L_\mathrm{X}$. \\
From $z$ and $\dot{M}_\star$ we calculate the expected inverse Compton X-ray luminosity according to equation (\ref{L_IC}), which however includes the normalization of the cosmic ray spectrum $Q_{\mathrm{p},0}$. This quantity depends on the fraction of supernova energy that goes into cosmic rays $\xi$ that is an open parameter of our model. We get an upper limit of $Q_{\mathrm{p},0}$ and accordingly $\xi$, from which we calculate the cosmic ray energy $E_\mathrm{CR}$, by the equalizing $L_\mathrm{IC}(Q_{\mathrm{p},0})$ with the observed X-ray luminosity of a galaxy $L_\mathrm{X,obs}$. In this step we imply that all the X-ray luminosity results from inverse Compton scattering. With the additional assumption of energy equipartition between cosmic rays and the magnetic field, an assumption which is commonly made in present-day galaxies \citep{BeckKrause2005}, we obtain an upper limit for the magnetic energy $E_\mathrm{mag}$. If one further estimates the volume of the galaxy, also an upper limit of the magnetic field strength $B$ is possible. \\
The single steps from above are described in more detail in the following. 

\subsection{Cosmic Ray Energy and the Equipartition Magnetic Field Strength}

\subsubsection{Cosmic Ray Energy as Function of the Free Parameter $\xi$}
\begin{figure}
  \includegraphics[width=0.5\textwidth]{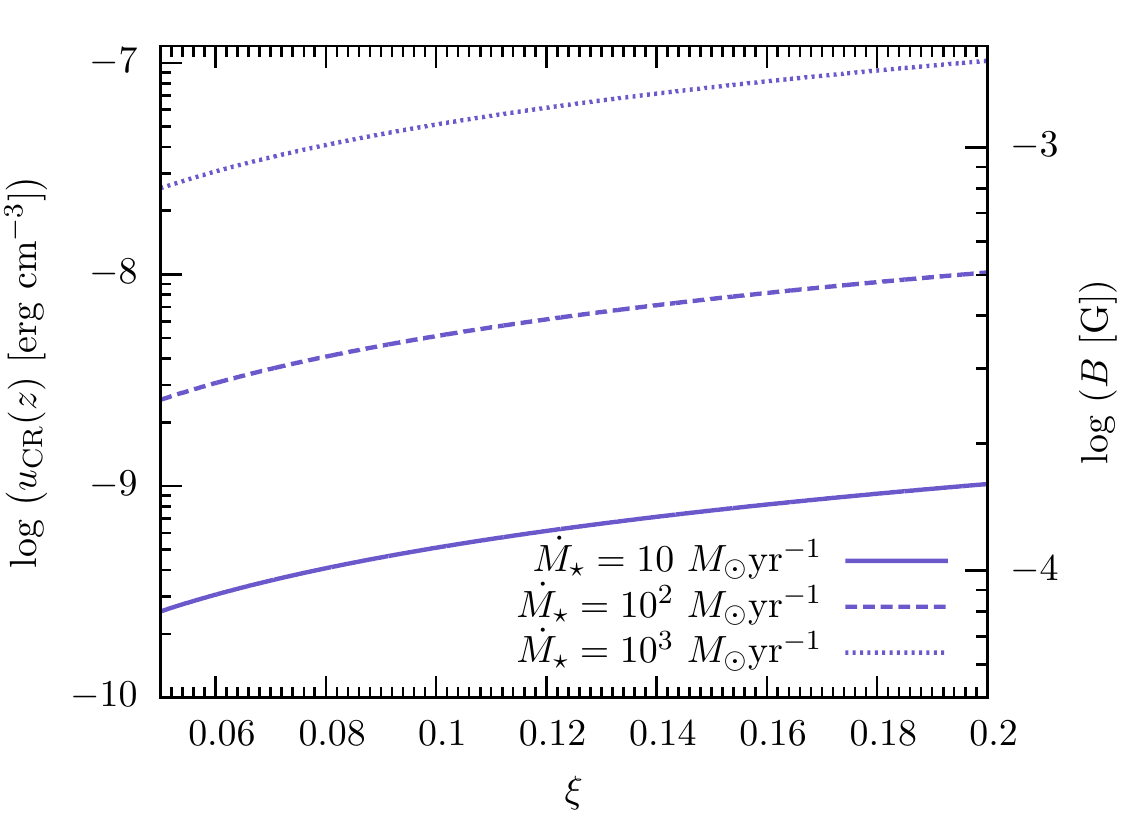}
  \caption{The cosmic ray density $u_\mathrm{CR}$ and the equipartition magnetic field strength $B$ as a function of the fraction of energy that goes into acceleration of cosmic rays in supernovae $\xi$. We use here a scaling of the galaxy volume proportional to $(1+z)^{-3}$. The different curves represent different star formation rates: $\dot{M}_\star = 10~\mathrm{M}_\odot \mathrm{yr}^{-1}$ (solid line), $\dot{M}_\star = 100~\mathrm{M}_\odot \mathrm{yr}^{-1}$ (dashed line) and $\dot{M}_\star = 1000~\mathrm{M}_\odot \mathrm{yr}^{-1}$ (dotted line). \vspace{0.3cm}}
\label{uCRB_xi}
\end{figure}

In section \ref{ExpectedIC} we determine the expected inverse Compton flux of a galaxy with a given cosmic ray spectrum. We model the cosmic ray spectrum by assuming that a fraction $\xi$ of the supernova energy is converted into kinetic energy of electrons and protons (see equation \ref{Qp0}). The energy density of cosmic rays $u_\mathrm{CR}$ as a function of the $\xi$ is 
\begin{eqnarray}
  u_\mathrm{CR}(\xi) = \frac{f_\pi~(m_\mathrm{p} c^2)^2~\gamma_\mathrm{p,0}^{2-\chi}~\tau_\pi(z)}{V(z)~(\chi -2)}~Q_{\mathrm{p},0}^\mathrm{p}(\xi),
\end{eqnarray}
with $Q_{\mathrm{p},0}^\mathrm{p}(\xi)$ given in equation (\ref{Np0}). We show $u_\mathrm{CR}(\xi)$ in figure \ref{uCRB_xi} for different fixed SFRs and a galaxy volume scaling as in equation (\ref{Vgal}). Note, that the redshift dependence of $u_\mathrm{CR}(\xi)$ cancels as the timescale of pion production is also proportional to $(1+z)^{-3}$. With increasing SFR the cosmic ray energy increases for a fixed $\xi$. This is intuitively clear.    \\

\subsubsection{Determination of the Cosmic Ray Energy from the Observed X-Ray Luminosity}
\begin{figure}
  \includegraphics[width=0.5\textwidth]{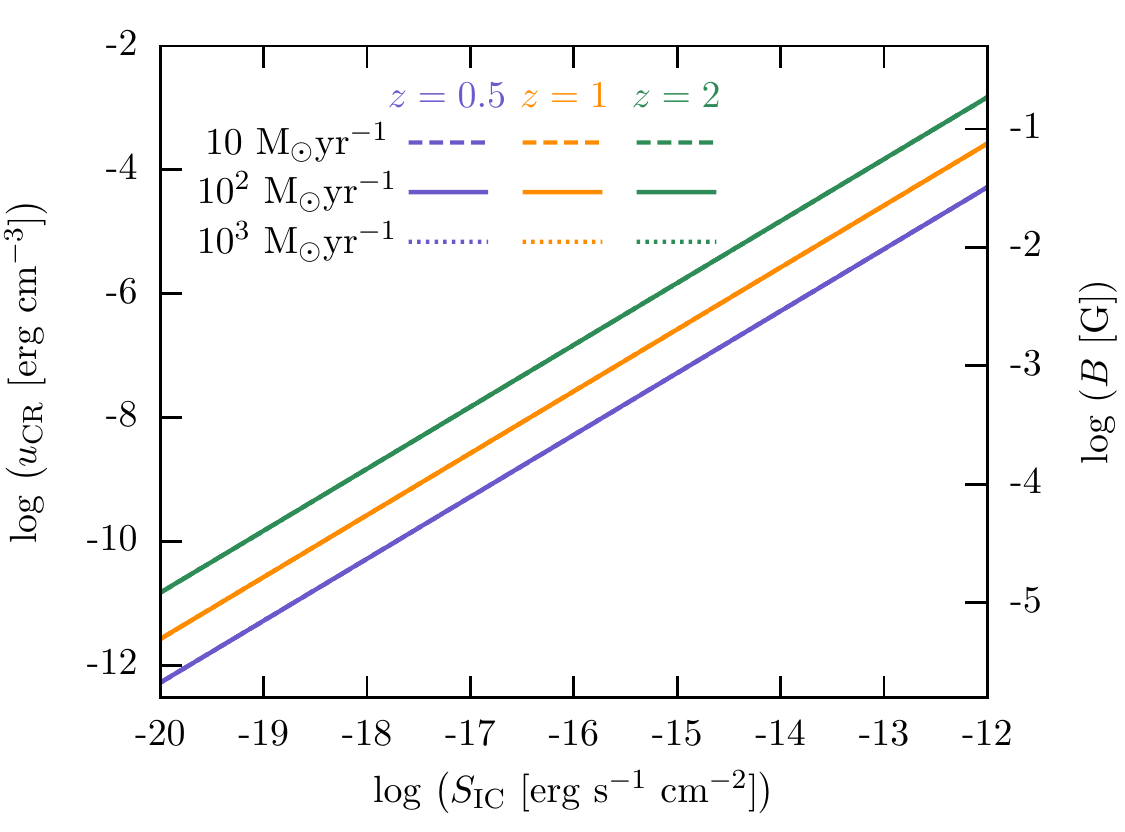}
  \caption{The energy density of cosmic rays (electrons + protons) $u_\mathrm{CR}$ as a function of the observed flux from inverse Compton scattering $S_\mathrm{IC}$. The red lines show the results at a redshift of $z=0.5$, the green lines at $z=1$ and the yellow lines at $z=2$. The different line styles refer to different star formation rates $\dot{M}_{\star}$. 
We use here the starburst model with different line styles corresponding to different star formation rates of $10~\mathrm{M}_\odot~\mathrm{yr}^{-1}$ (dashed lines), $100~\mathrm{M}_\odot~\mathrm{yr}^{-1}$ (solid lines) and $1000~\mathrm{M}_\odot~\mathrm{yr}^{-1}$ (dotted lines). Note that there is no difference between the models with different SFR. On the right hand side y axes we show the corresponding equipartition magnetic field strength $B$. \vspace{0.3cm}}
\label{uCRB_S__givenSFR}
\end{figure}

From the data we will now directly determine the normalization $Q_{\mathrm{p},0}$ and thus the free parameter $\xi$, to which the X-ray luminosity from inverse Compton scattering (\ref{L_IC}) is directly proportional. Solving equation (\ref{L_IC}) for $Q_{\mathrm{p},0}$ and using the observed luminosity $L_\mathrm{X,obs}$ as an input yields:
\begin{eqnarray}
  Q_{\mathrm{p},0}(L_\mathrm{X,obs}) & = & \frac{f_\mathrm{sec} h^{2+\chi/2} m_\mathrm{e}^{\chi-2} \sigma_\mathrm{T} (\chi-2)}{16 c k^{3+\chi/2} m_\mathrm{p}^{\chi} F(\chi) f_\pi \pi^2 r_0^2} \frac{u_\mathrm{ISRF}}{(1+z)^{\chi/2-1}} \nonumber \\
     &  & \left(\sum_i f_i~T_i^{3 + \chi/2}\right)^{-1} \frac{L_\mathrm{X,obs}}{(\nu_1^{1-\chi/2} - \nu_2^{1-\chi/2})}. \nonumber \\
     &  &
\label{Ngamma0pLx}
\end{eqnarray}
Note, that (\ref{Ngamma0pLx}) depends on the star formation rate only via the total energy density of the radiation field $u_\mathrm{ISRF} \propto \sum_i f_i~T_i^4$ and the sum over $f_i~T_i^{3 + \chi/2}$. With a value of $\chi$ very close to 2, these two terms cancel and $u_\mathrm{CR}$ becomes almost independent of the star formation rate. Also the dependence on redshift is small as again $(1+z)^{\chi/2-1}$ is almost constant. \\
With equation (\ref{Ngamma0pLx}) our free parameter, the energy input of supernovae into cosmic ray acceleration $\xi$, can be expressed as
\begin{eqnarray}
  \xi(L_\mathrm{X,obs}) = \frac{(m_\mathrm{p} c^2)^2 ~ \gamma_\mathrm{p,0}^{2-\chi}}{\xi E_\mathrm{SN} \dot{N}_\mathrm{SN} (\chi-2) f_\pi \tau_\mathrm{\pi}} Q_{\mathrm{p},0}(L_\mathrm{X,obs}) ,
\end{eqnarray}
where we used equation (\ref{Np0}). Further, the energy density of cosmic rays as a function of the observed X-ray luminosity can then be calculated with
\begin{eqnarray}
  u_\mathrm{CR}(L_\mathrm{X,obs})  =  Q_{\mathrm{p},0} (L_\mathrm{X,obs}) \frac{\gamma_\mathrm{p,0}^{2-\chi} (m_\mathrm{p} c^2)^2}{\chi-2} f_\pi \frac{\tau_\pi(z)}{V_\mathrm{gal}(z)}. \nonumber \\  
\end{eqnarray}
The luminosity is converted into the observed flux by equation (\ref{SICnu}). We plot the cosmic ray energy against the flux $S_\mathrm{IC}$ in figure \ref{uCRB_S__givenSFR}. The different line colors represent different redshifts, while the line styles indicate different SFRs. The figure shows clearly that the SFR dependence vanishes in our model of the inverse Compton scattering, as the individual lines match almost perfectly. Note, however, that the flux itself depends on the SFR! As shown in figure \ref{SIC_SFR}, the flux increases with $\dot{M}_\star$. Actually, equations (\ref{SICnu}), (\ref{L_IC}), (\ref{Np0}), and (\ref{SNrate}) indicate that $S_\mathrm{IC} \propto Q_{\mathrm{p},0} \propto \dot{N}_\mathrm{SN} \propto \dot{M}_\star$, connecting the inverse Compton flux with the number of cosmic rays, which is in our model directly proportional to the supernova rate and thus to the SFR. 

\subsubsection{Equipartition Assumption}
In present-day galaxies the different energy components of the ISM, like the kinetic energy of the gas, the cosmic rays and the magnetic fields, are almost in equilibrium. This is a result of the dynamical interplay of the individual components and known as energy equipartition. It provides a very important tool for studying the magnetic field strength from the observed synchrotron radiation, which is emitted by cosmic ray electron traveling through the magnetized interstellar medium \citep{BeckKrause2005}. \\
In highly redshifted galaxies magnetic fields were assumed to be unimportant, because the timescales of a galactic large-scale dynamo are high and thus no strong magnetic field can result from this mechanism. However, recent semi-analytical \citep{SchoberEtAl2013} and numerical simulations \citep{BeckEtAl2012,LatifEtAl2013} have shown that a turbulent dynamo can actually amplify weak magnetic seed fields in galaxies on Myr timescales by converting turbulent kinetic energy into magnetic energy. The turbulent dynamo can amplify fields up to a certain fraction of the turbulent kinetic energy even on galactic length scales (see, e.g., \citet{FederrathEtAl2011.2}). With this strong unordered magnetic fields we can again use the assumption of energy equipartition:
\begin{equation}
  u_\mathrm{CR} = u_\mathrm{B}.
\end{equation}
With the magnetic energy density, $u_\mathrm{B} = B^2/(8\pi)$, the magnetic field strength can be calculated as
\begin{equation}
  B = \left(8 \pi u_\mathrm{CR}\right)^{1/2}.
\end{equation}
Putting the equations together, one can show that the resulting magnetic field $B$ scales with $Q_{\mathrm{p},0}^{1/2}$ and also depends on the slope of the cosmic ray spectrum $\chi$. The equipartition field strengths are given on the right side of figures \ref{uCRB_xi} and \ref{uCRB_S__givenSFR}.

\subsection{Test Case M82}
\label{TestM82}

As a test case for our model we choose the local starburst galaxy M82, for which various observations of cosmic rays and magnetic fields are available. The basic properties of the starburst core of M82, like the spacial extension, the density, and the SFR, are provided in table \ref{Table_Props}. \\
Observations show, that the total (unabsorbed) X-ray luminosity from M82 is $1.6\times10^{41}$ erg s$^{-1}$ in the 0.3-10 keV range and $3\times10^{40}$ erg s$^{-1}$ in the hard 2-10 keV range \citep{MoranLehnert1997}. Assuming that all the luminosity results from inverse Compton scattering with our model we get for the fraction of energy in cosmic rays $\xi=1.41$ in the 0.3-10 keV range and $\xi=0.60$ in the hard X-ray regime. This yields $u_\mathrm{CR} = 8.51\times10^{-9}~\mathrm{erg~cm}^{-3}$ and $B = 4.62 \times 10^{-4}~\mathrm{G}$ and $u_\mathrm{CR} = 3.84\times10^{-9}~\mathrm{erg~cm}^{-3}$ and $B = 3.11 \times 10^{-4}~\mathrm{G}$ in total and the hard X-ray regime. \\
From the calculated value of the fraction of supernova energy that goes into cosmic rays $\xi$, one already sees that we are overestimating the inverse Compton luminosity and thus the cosmic ray energy. A natural upper limit of $\xi$ is 1, when the entire supernova energy goes into cosmic ray acceleration. Reasonable values of $\xi$, however, lie between 0.05 and 0.2. Assuming $\xi=0.1$, which is our fiducial value, we see that the magnetic energy is overestimated with our model by a factor of 14 or 6, respectively. Consequently the magnetic field strength, which depends on the square root of the magnetic energy, is overestimated by a factor of 3.7 and 2.4, respectively. Thus, there are other important components contributing to the total X-ray luminosity of M82, like X-ray binaries. Without modeling these contributions in detail, we can only get upper limits for the cosmic ray density and equilibrium magnetic field strength. \\
Detailed models of starburst core of M82 yield in comparison with gamma ray observations a magnetic field strength of $2.50\times10^{-4}$ G \citep{Yoast-HullEtAl2013}. Other authors find different values for the field strength ranging from $2.0 \times10^{-4} - 1.0 \times10^{-3}$ G \citep{PaglioneAbrahams2012} to $1.2 - 2.9 \times10^{-4}$ G \citep{deCeadelPozoEtAl2009}, while the typical number densities in the models also deviate from each other. Thus, the order of magnitude of the magnetic field is similar to our results and we are overestimating the magnetic field strength only by a factor below 2. This leads to the conclusion that inverse Compton scattering is one of the dominant X-ray sources in the starburst region of M82.

\subsection{Available Observational Data at High Redshifts}
\label{DataSet}

\begin{table}
\centering
     \begin{tabular}{ccccc}
      \hline  \hline   
      \tabh   ALESS ID 	& $z$   & $\dot{M}_\star$               & $L_\mathrm{X,obs}$   \\
      \tabh   ~        	& ~     & $[\mathrm{M}_\odot~\mathrm{yr}^{-1}]$  & $[\mathrm{erg~s}^{-1}]$ \\
      \hline   
      \tab 011.1 	& 2.68 	& 789 				& $3.16\times 10^{43}$ \\
      \tab 017.1 	& 2.04 	& 161 				& $2.51\times 10^{42}$ \\
      \tab 045.1 	& 2.34 	& 350 				& $1.58\times 10^{42}$ \\
      \tab 057.1 	& 2.94	& 439				& $5.01\times 10^{43}$ \\
      \tab 066.1 	& 1.31 	& 322 				& $3.16\times 10^{44}$ \\
      \tab 067.1 	& 2.12 	& 528 				& $2.51\times 10^{42}$ \\
      \tab 070.1	& 2.33 	& 789	 			& $1.58\times 10^{43}$ \\
      \tab 073.1	& 4.76 	& 556				& $5.01\times 10^{43}$ \\
      \tab 084.1 	& 2.26 	& 267 				& $1.00\times 10^{43}$ \\
      \tab 114.2 	& 1.61	& 261 				& $6.31\times 10^{42}$ \\
      \hline  \hline 
    \end{tabular}
\caption{The properties of the ALESS sub-mm objects with X-ray counterparts in the E-CDF-S, which were discovered by \cite{WangEtAl2013}. We list here the ALESS ID, the redshift $z$, the star formation rate $\dot{M}_\star$ in $\mathrm{M}_\odot~\mathrm{yr}^{-1}$ (converted into Kroupa IMF) and the non-corrected X-ray luminosity $L_\mathrm{X,obs}$ in $\mathrm{erg~s}^{-1}$. For more details see section \ref{DataSet}. }
  \label{TableObsInput}
\end{table}
\begin{figure}
  \includegraphics[width=0.5\textwidth]{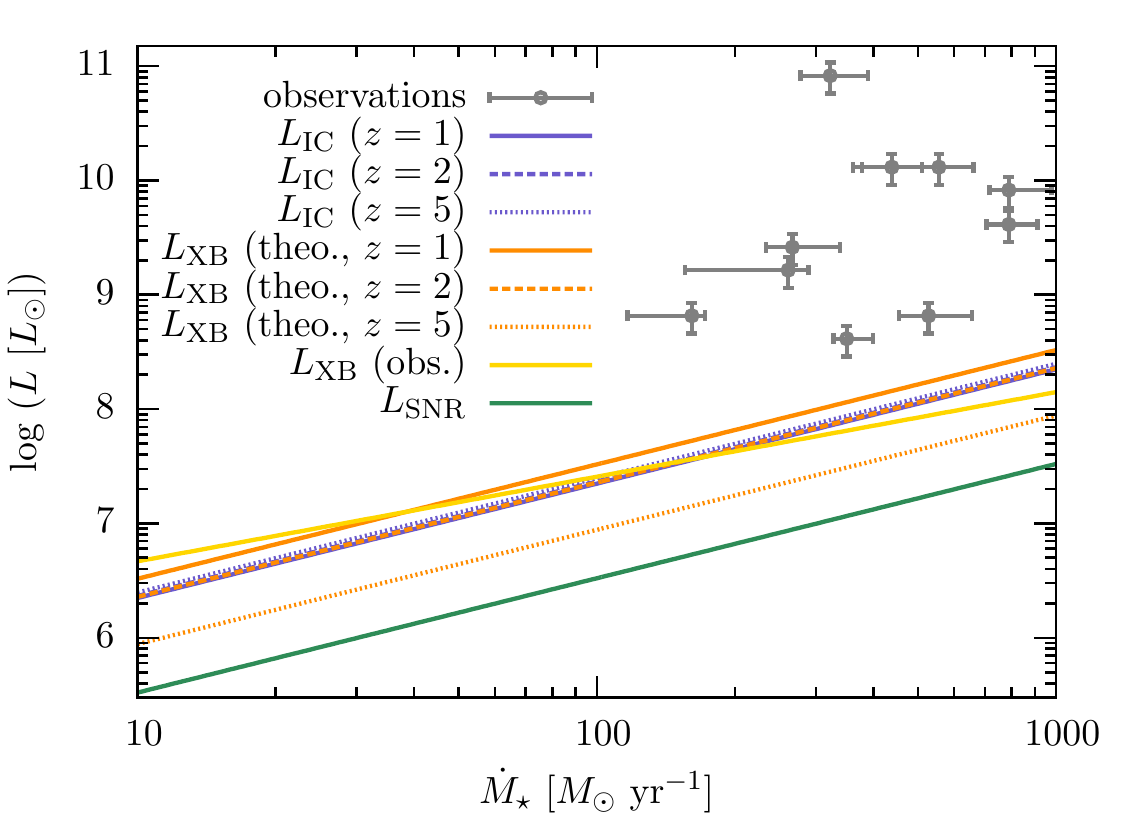}
  \caption{The 0.5-8 keV X-ray luminosity as a function of star formation rate $\dot{M}_\star$. We compare the inverse Compton luminosity $L_\mathrm{IC}$ with the luminosity of X-ray binaries $L_\mathrm{XB}$ and the one of supernova remnants. Also plotted are the X-ray luminosities (with an 30\% error as discussed in \citet{XueEtAl2011}) of the \textit{Chandra} deep field galaxies from \citet{WangEtAl2013}. Different line styles represent different redshifts in the model from $z=1$ (solid lines) to $z=2$ (dashed lines) to $z=5$ (dotted lines). \vspace{0.3cm}}
  \label{L_SFR}
\end{figure} 
\begin{table*} 
\centering
 \begin{tabular}{l ccccc ccc}
  \hline  \hline  
  \tabh ~ 					& ~& ~			& ALESS ID 045.1  	& ~			& ~& ~			& ALESS ID 067.1  	& ~  \\

  \tabh    ~     			   	& ~& $\chi=2.1$    	& $\chi=2.2$   		& $\chi=2.3$ 		& ~& $\chi=2.1$  		& $\chi=2.2$   		& $\chi=2.3$ 	 \\
  \hline   
  \tabh  $\xi$    			   	& ~& 0.893		  	& 0.501  		& 0.371			& ~& 0.944	 		& 0.532			& 0.395	        \\  
  \tabh  $U_\mathrm{CR}$ [erg]   	   	& ~& $7.12\times10^{54}$	& $3.99\times10^{54}$ 	& $2.96\times10^{54}$	& ~& $1.39\times10^{55}$	& $7.84\times10^{54}$ 	& $5.82\times10^{54}$ 	     \\  
  \hline
  \tabh  $V_\mathrm{gal} = 0.1~V_\mathrm{M82}$  & ~&    	~	  	&   	~		&  	~	 	& ~& 	 ~		& 	~		&        	 \\  
  \tabh  $u_\mathrm{CR}$ [erg~cm$^{-3}$] 	& ~& $4.28\times10^{-8}$  	& $2.40\times10^{-8}$ 	& $1.78\times10^{-8}$	& ~& $8.37\times10^{-8}$  	& $4.72\times10^{-8}$ 	& $3.50\times10^{-8}$       \\  
  \tabh  $B$ [G] 				& ~& $1.04\times10^{-3}$	& $7.77\times10^{-4}$ 	& $6.69\times10^{-4}$ 	& ~& $1.45\times10^{-3}$	& $1.09\times10^{-3}$ 	& $9.38\times10^{-4}$  \\ 
  \hline
  \tabh  $V_\mathrm{gal} = V_\mathrm{M82}$  	& ~&    	~	 	&   	~	 	&  	~		& ~& 	 ~		& 	~		&       	   \\  
  \tabh  $u_\mathrm{CR}$ [erg~cm$^{-3}$]	& ~& $4.28\times10^{-9}$  	& $2.40\times10^{-9}$ 	& $1.78\times10^{-9}$ 	& ~& $8.37\times10^{-9}$  	& $4.72\times10^{-9}$ 	& $3.50\times10^{-9}$ 	 \\  
  \tabh  $B$ [G] 				& ~& $3.28\times10^{-4}$	& $2.46\times10^{-4}$ 	& $2.11\times10^{-4}$ 	& ~& $4.59\times10^{-4}$	& $3.44\times10^{-4}$ 	& $2.97\times10^{-4}$ 	      \\ 
  \hline
  \tabh  $V_\mathrm{gal} = 10~V_\mathrm{M82}$ 	& ~&    	~		&   	~		&  	~		& ~& ~			& ~ 			& ~     	 \\  
  \tabh  $u_\mathrm{CR}$ [erg~cm$^{-3}$] 	& ~& $4.28\times10^{-10}$ 	& $2.40\times10^{-10}$ 	& $1.78\times10^{-10}$  & ~& $8.37\times10^{-10}$ 	& $4.72\times10^{-10}$ 	& $3.50\times10^{-10}$         \\  
  \tabh  $B$ [G] 				& ~& $1.04\times10^{-4}$	& $7.77\times10^{-5}$ 	& $6.69\times10^{-5}$ 	& ~& $1.45\times10^{-4}$	& $1.09\times10^{-4}$ 	& $9.38\times10^{-5}$ 	        \\ 
  \hline  \hline  
    \end{tabular}
\caption{Upper limits on the cosmic ray energy, the cosmic ray density and the magnetic field strength for two objects of the data catalog of \cite{WangEtAl2013} ALESS ID 045.1 and ALESS ID 067.1, which have not been identified as hosting an AGN. We present results for the assumption that all X-ray luminosity is produced in inverse Compton scattering. As there are other processes emitting X-rays, this assumption results only in upper limits, which reflects in the high values of the fraction of supernova energy that is transformed into kinetic energy of cosmic rays $\xi$ (theoretically expected: $\xi\approx 0.1$, see text). We show the resulting value of $\xi$ obtained with this assumption in the first line. In the second line we present the total cosmic ray energy. For determining the cosmic ray densities $u_\mathrm{CR}$ and the equipartition field strengths $B$ we use three different volumes of the starburst region: the volume of the starburst region in M82 $V_\mathrm{M82}$ and volumes ten times smaller and larger then $V_\mathrm{M82}$. \vspace{0.3cm}}
 \label{TableResults}
\end{table*}
The combination of X-ray data of the \textit{Chandra} deep fields with the sub-millimeter data from ALMA, provides new insides in the properties of distant galaxies. Our analyses is based on the processed data given in \citet{WangEtAl2013}, who have identified 10 sub-mm counter sources with objects in the \textit{Chandra} Deep Field - South (CDF-S). \\
\citet{WangEtAl2013} use the X-ray data from the 4 Ms CDF-S \citep{LehmerEtAl2005} and the 250 ks E-CDF-S survey \citep{XueEtAl2011}. The E-CDF-S has been observed by an ALMA Cycle 0 survey at 870 $\mu$m, which is called the ALMA LABOCA E-CDF-S Submm Survey (ALESS) \citep{HodgeEtAl2013, KarimEtAl2013}. The whole ALESS survey detected 99 sub-mm galaxies (SMGs), 10 of which \citet{WangEtAl2013} could identify with X-ray counterparts. These SMGs are listed in table \ref{TableObsInput}, where besides their X-ray luminosities $L_\mathrm{X,obs}$, the corresponding redshifts $z$ and SFRs $\dot{M}_\star$ are provided. Note, that \citet{WangEtAl2013} used a Salpeter IMF to determine the SFRs. We converted their values to a Kroupa IMF by dividing by a factor of 1.8. Further, we use here the luminosity values, which are not corrected for dust attenuation, as we assume that the latter is most efficient for the X-ray emission from the central black hole. Most of the redshifts $z$ are observed spectroscopically within the zLESS survey (Danielson et al.~2013, in preparation) or taken from literature, expect for ALESS 45.1, which is observed photometrically \citep{SimpsonEtAl2013}. The SFRs listed in table \ref{TableObsInput} are derived by \cite{WangEtAl2013} from the correlation with the infrared luminosity by \citet{Kennicutt1998} (see equation \ref{SFR_LIR}). \\
The observed X-ray luminosities from \cite{WangEtAl2013} are shown as a function of SFR in figure \ref{L_SFR}. All the sources from the catalog have extremely high SFRs from 161 $\mathrm{M}_\odot\mathrm{yr}^{-1}$ up to 789 $\mathrm{M}_\odot\mathrm{yr}^{-1}$. In the figure we show also the expected luminosity of X-ray binaries from the analytical model of \citet{GhoshWhite2001} and the observational correlation from \citet{LehmerEtAl2010} $L_\mathrm{XB}$, as well as the luminosity from supernova remnants $L_\mathrm{SNR}$ and the inverse Compton luminosity $L_\mathrm{IC}$. Most of the observed galaxies have very high luminosities, even above the X-ray binary predictions. Due to this fact and also as a result of additional tests most of the galaxies have clearly been classified as AGN hosts, except for ALESS 045.1 and ALESS 067.1. The X-ray luminosity of these two galaxies could be explained by X-ray binary emission. Alternatively a huge contribution of the luminosity could come from the inverse Compton scattering. \\
In the following we use ALESS 045.1 and ALESS 067.1 as examples, for which we will derive upper limits for the cosmic ray density from the inverse Compton effect. We make the assumption that the entire X-ray luminosity observed comes from inverse Compton scattering. With our model we then result in a value for the free parameter $\xi$, which is given in the first line of table \ref{TableResults} for the two galaxies and different slopes of the cosmic ray spectrum $\chi$. Note, that all the calculated values of $\xi$ are larger than our fiducial value $\xi=0.1$. This indicates that we are overestimating the contribution of the inverse Compton luminosity in all cases by factors up to 9. Consequently the magnetic field strength will be overestimated by a factor of roughly 3. The cosmic ray energies given in the second line of \ref{TableResults} can thus be only treated as upper limits. In the following lines of the table we present the cosmic ray densities and the equipartition magnetic field strengths for different fixed galaxy radii. As the galaxies are not spatially resolved in the observations, we do not have any information about their radius. With radii comparable to the one of M82 plus radii ten times smaller and larger then that (see numbers listed in table \ref{Table_Props}), we get hints to the energy density in the galaxies. For ALESS 045.1, which has a SFR of $350~\mathrm{M}_\odot \mathrm{yr}^{-1}$, we find values between $7.77\times10^{-4}$ G and $7.77\times10^{-5}$ G for our fiducial cosmic ray spectrum with $\chi=2.2$. For ALESS 067.1 the upper limits for the magnetic field strength is slightly higher with $B\approx1.09\times10^{-4}$ G to $B\approx1.09\times10^{-5}$ G.

\subsection{Uncertainties in the Model and Possible Extensions}

Our model for the determination of cosmic ray densities and magnetic fields in galaxies includes many assumptions. We will discuss the individual problems in the following starting with effects that lead to an overestimation of the final results. \\
A caveat in our calculation remains the influence of additional X-ray sources, such as X-ray binaries and AGNs. The contribution of these can potentially be investigated in more detail with observations in additional energy ranges. Furthermore detailed models, especially of the luminosity from the accretion on the central supermassive black hole are required. Galaxies without AGNs would be easier to handle in the framework of our model and result in better estimates of the cosmic ray properties. It is, however, very hard to find such objects at high redshifts, as their total luminosity is very low. Hopefully, future observatories like \textit{Athena+} will detect more starburst galaxies without AGNs at far distances. \\
For the energy loss of cosmic ray electrons to be dominated by inverse Compton scattering one needs to find galaxies at high redshift and with large star formation rates (see figure \ref{timescales_SFR} and the discussion in section \ref{Breakdown}). If inverse Compton scattering is not the most important loss channel, but similar or below synchrotron emission or bremsstrahlung, the model of the steady state spectrum needs to be modified. A detailed analysis of the different energy loss mechanisms will be an interesting future test.

\section{Conclusions}
\label{Conclusions}
In this work we construct a model for the X-ray emission of star-forming galaxies via inverse Compton scattering as a function of redshift. We model the star formation rate (SFR) history, the evolution of the interstellar radiation field (ISRF) and the cosmic ray spectrum. The inverse Compton scattering process between high energy cosmic ray electrons and the ISRF is quantified and analyzed in terms of different properties of the galaxy. We focus on two galaxy models: a galaxy with normal star formation rate, similar to the Milky Way, and a starburst galaxy similar to M82. With a detailed description of the ISRF and the steady state cosmic ray spectrum we are able to calculate the expected inverse Compton luminosity. \\
In order to estimate the significance of the inverse Compton scattering compared to other galactic X-ray sources, we investigate the role of X-ray binaries, which are one of the main X-ray sources in nearby galaxies. We summarize an analytical model for the number of high-mass X-ray binaries (HMXB) and low-mass X-ray binaries (LMXB) by \citet{GhoshWhite2001}. For comparison we also discuss an observational correlation for X-ray binary luminosity by \citet{LehmerEtAl2010}. Furthermore, we estimate the influence of supernova remnants on the total galactic X-ray luminosity. \\
In the last part of the paper (section \ref{Examples}) we apply our model to real observations. As observational input we use M82 as a test case and two higher redshifted galaxies of the data set of \citet{WangEtAl2013}, which have not been identified as hosts of active galactic nuclei (AGNs). We compare the observed X-ray luminosity with the one resulting from our inverse Compton model. This way we can fix the free parameter in our model, namely the normalization of the cosmic ray spectrum. In the next step we calculate the total energy of cosmic rays and assume that it is in equipartition with the magnetic energy. \\
The main findings of this work are:
\begin{itemize}
\renewcommand{\labelitemi}{$\bullet$}
\item{The spectral energy distribution $u_{\mathrm{ISRF},\nu}$ of a normal galaxy is dominated by the cosmic microwave background (CMB), while the one of a starburst galaxy is dominated by the cold infrared (IR) component at least at moderate redshifts (see figure \ref{ISRF_nu}). The strong IR component makes inverse Compton scattering in starburst galaxies more efficient (see also figures \ref{Efficiency_nuin} and \ref{L_z__z}).}
\item{Our analysis of the energy loss timescales of cosmic ray electrons (see figure \ref{timescales_SFR}) has shown, that the inverse Compton scattering is not dominant in galaxies with normal star formation. At low redshifts $z$ bremsstrahlung and synchrotron emission are most important. With increasing redshift the inverse Compton timescale decreases, but even at $z=5$, bremsstrahlung is still dominating. On the other hand in starburst galaxies energy losses proceed mostly via inverse Compton scattering. These galaxies are thus in the focus of this work.}
\item{The X-ray flux from pure inverse Compton scattering can be detected with \textit{Chandra} up to $z\approx1$ for starburst galaxies with $\dot{M}_\star \gtrsim 200~\mathrm{M}_\odot \mathrm{yr}^{-1}$. With the future X-ray observatory \textit{Athena+} detections up to $z\gtrsim2$ will be possible (see figure \ref{SIC_SFR}).}
\item{Comparison of the expected inverse Compton luminosity with other X-ray sources shows that supernova remnants are negligible. X-ray binaries play a more important role. In our model their luminosity is a factor of 2 brighter than the inverse Compton luminosity at present day. At redshifts above roughly 2 inverse Compton luminosity becomes comparable or even dominant over the X-ray binaries (see figure \ref{LXB_z}).}
\item{With our model the energy density of cosmic rays can be determined directly from the observed X-ray flux under the assumption that the flux only origins from inverse Compton scattering. The results for different redshifts are plotted in figure \ref{uCRB_S__givenSFR}.}
\item{We apply our model to the two galaxies from the data set of \citet{WangEtAl2013} that have not been clearly identified as hosting an AGN. Our results for the fraction of energy going from supernovae into cosmic ray acceleration is higher then the theoretically expected value of 10 percent. This suggests that we are overestimating the inverse Compton luminosity and thus the energy of cosmic rays by a factor of up to 9 (see discussion in section \ref{TestM82}). Our results for the cosmic ray density and the equipartition field strengths are thus only upper limits. Depending on the galactic volume we find values for the magnetic field strength of roughly $10^{-4}-10^{-3}$ G for the exemplary galaxies (see table \ref{TableResults}).}
\end{itemize}
There are several uncertainties in our model including the modeling of the cosmic ray spectrum, the additional X-ray sources and the evolution and total size of the galaxy volume. Most of the galaxies observed at high redshift include AGNs. We try not to include the X-ray emission of these by using the uncorrected X-ray luminosities given in \citet{WangEtAl2013}. However, we still substantially overestimate the X-ray luminosity from inverse Compton scattering. It thus is essential to model the X-ray emission of galaxies in more detail in future. \\
We expect that also the available data of high redshifted galaxies will increase in the next years. For our studies especially observations of distant starburst galaxies without active galactic nuclei would be important. With X-ray data from the \textit{Chandra} deep fields the next step would be to identify infrared counterparts of X-ray galaxies, in order to determine their SFR. This is possible with the \textit{ALMA} telescope. Further the next generation of X-ray telescopes is planned and we hopefully will receive a lot of data with \textit{Athena+}. \\
With these new technologies our knowledge of the origin and evolution of galactic magnetic field hopefully will increase. This will help us to understand moreover the evolution of galaxies in total, as magnetic fields play a crucial role in many physical processes in the interstellar medium and the dynamics of the whole galaxy.

\section*{Acknowledgments}

We acknowledge fruitful discussions with Simon Glover and Fabian Walter. Further we thank for funding through the {\em Deutsche Forschungsgemeinschaft} (DFG) in the {\em Schwer\-punkt\-programm} SPP 1573 ``Physics of the Interstellar Medium'' under grant KL 1358/14-1 and SCHL 1964/1-1. This work has also been financially supported by the {\em Baden-W\"urttemberg-Stiftung} via contract research (grant P-LS-SPII/18) in their program ``Internationale Spitzenforschung II'' as well as the DFG via the SFB 881 ``The Milky Way System'' in the sub-projects B1 and B2. J.~S.~acknowledges the support by IMPRS HD. D.~R.~G.~S.~ further thanks for funding via the SFB 963/1 (project A12) on ``Astrophysical flow instabilities and turbulence''. R.S.K.~acknowledges support from the European Research Council under the European Community's Seventh Framework Programme (FP7/2007-2013) via the ERC Advanced Grant STARLIGHT (project number 339177).


\end{document}